\documentclass[useAMS,usenatbib]{mn2e}
\usepackage{aalongtable}

\usepackage{graphicx}

\usepackage{lscape}
\usepackage{graphicx}
\usepackage{txfonts}
\newcommand{\cz}{\ensuremath{C_z}}
\newcommand{\Bz}{\ensuremath{\langle B_z\rangle}}
\def\gtrsim{\mathrel{\hbox{\rlap{\hbox{\lower4pt\hbox{$\sim$}}}\hbox{$>$}}}}

\def\ltsim{\mathrel{\hbox{\rlap{\hbox{\lower4pt\hbox{$\sim$}}}\hbox{$<$}}}}

\def\kms{\hbox{{\rm km}\,{\rm s}$^{-1}$}}

\def\bz{\hbox{$\langle B_z\rangle$}}

\title[A search for strong, ordered magnetic fields in Herbig Ae/Be stars]{A search for strong, ordered magnetic fields in Herbig Ae/Be stars\thanks{Based on observations from the ESO telescopes at the La Silla Paranal Observatory under programme ID 072.C-0447, DDT-272.C-5063, 074.C-0442.}}

   \author[G.A. Wade, S. Bagnulo, D. Drouin, J.D. Landstreet, D. Monin]{G.A. Wade$^{1}$, S. Bagnulo$^{2}$, D. Drouin$^{1}$, J.D. Landstreet$^{3}$, D. Monin$^{4}$\\
   $^1$Department of Physics, Royal Military College of Canada, 
   PO Box 17000, Station 'Forces', Kingston, Ontario, Canada K7K 4B4\\
   $^2$European Southern Observatory, Casilla 19001, Santiago 19, Chile\\
   $^3$Department of Physics \& Astronomy, The University of Western Ontario, London, Ontario, Canada, N6A 3K7\\
   $^4$National Research Council of Canada, Herzberg Institute of Astrophysics, 5071 West Saanich Rd., Victoria, BC, Canada, V9E 2E7}

\begin{document}                                                             



\maketitle

\begin{abstract}
{The origin of magnetic fields in intermediate-mass and high-mass stars is fundamentally a mystery. Clues toward solving this basic astrophysical problem can likely be found at the pre-main sequence (PMS) evolutionary stage.}
{With this work, we perform the largest and most sensitive search for magnetic fields in pre-main sequence Herbig Ae/Be (HAeBe) stars. We seek to determine whether strong, ordered magnetic fields, similar to those of main sequence Ap/Bp stars, can be detected in these objects, and if so, to determine the intensities, geometrical characteristics, and statistical incidence of such fields.}
{Sixty-eight observations of 50 HAeBe stars have been obtained in circularly polarised light using the FORS1 spectropolarimeter at the ESO VLT.
}{An analysis of both Balmer and metallic lines reveals the possible presence of weak longitudinal magnetic fields in photospheric lines of two HAeBe stars, HD 101412 and BF Ori. Results for two additional stars, CPD-53 295 and HD 36112, are suggestive of the presence of magnetic fields, but no firm conclusions can be drawn based on the available data. The intensity of the {longitudinal} \rm fields detected in HD 101412 and BF Ori suggest that they correspond to globally-ordered magnetic fields with surface intensities of order 1~kG. On the other hand, no magnetic field is detected in {4} other HAeBe stars in our sample in which magnetic fields had previously been confirmed. {Monte Carlo simulations of the longitudinal field measurements of the undetected stars allow us to place an upper limits of about 300~G on the general presence of aligned magnetic dipole magnetic fields, and of about 500~G on perpendicular dipole fields.} Taking into account the results of our survey and other published results, we find that the observed bulk incidence of magnetic HAeBe stars in our sample is 8-12\%, in good agreement with that of magnetic main sequence stars of similar masses. We also find that the rms longitudinal field intensity of magnetically-detected HAeBe stars is about 200~G, similar to that of Ap stars and consistent with magnetic flux conservation during stellar evolution. These results {are all in agreement with} the hypothesis that the magnetic fields of main sequence Ap/Bp stars are fossils, which already exist within the stars at the pre-main sequence stage. {Finally, we explore the ability of our new magnetic data to constrain magnetospheric accretion in Herbig Ae/Be stars, showing that our magnetic data are not consistent with the general occurrence in HAeBe stars of magnetospheric accretion as described by the theories of K\"onigl (1991) and Shu et al. (1994).}}
\end{abstract}


\section{Introduction}

About 5-10\% (e.g. Wolff 1968) of all intermediate mass (B and A) main sequence stars exhibit
organised magnetic fields with strengths ranging between a few
hundred and a few tens of thousands of gauss. The
presence of these fields has important consequences for the structure of
the atmospheres of these Ap/Bp stars, suppressing { large-scale mixing} and
leading to an amazing array of atmospheric peculiarities which
effectively define these stars spectroscopically (e.g. Adelman 1993). 

Their magnetic fields have clearly defined observational and physical
properties which are totally different from those of the sun and
other late-type stars. The fields are strong, with important dipolar
components and filling factors near 100\%; the field
structure is static in the rotating stellar reference frame, and appears
to be stable on timescales of decades; and the
fields show only weak or no correlation with projected rotational
velocity, main sequence age, and other physical parameters. 

Because these characteristics contrast so strongly with those of active cool
stars (the only other broad class of non-degenerate stars in which
magnetic fields can be studied systematically), the magnetic Ap/Bp stars
represent an important laboratory for investigating the fundamental physics
associated with stellar and cosmic magnetism. { In particular, because stars in this mass range are progenitors of both white dwarfs and neutron stars, their magnetic characteristics may provide important clues needed in order to understand magnetism in these degenerate objects. }

Remarkably, very little is known about the formation and evolution of
these magnetic fields.  Recent investigations by Bagnulo et al. (2003, 2004, 2006) 
and P\"ohnl et al. (2003, 2005), showing that magnetic and chemically
peculiar A-type stars are present in very young open clusters, strongly support
the view that the {\em fields are present in these stars at the time they
reach the main sequence.} This suggests that the magnetic fields observed in main sequence Ap/Bp stars are present, and possibly produced, within their
pre-main sequence (PMS) progenitors. 
This idea is also supported by the theoretical work of St\c {e}pie\'n
(2000) and of St\c {e}pie\'n and Landstreet (2002), which suggest that strong magnetic fields are required at the pre-main sequence phase in order to produce the slow rotation which is commonly observed in the Ap/Bp stars.
 The primary aim of this study is to test these propositions by
searching for direct evidence of magnetic fields in the photospheres of a statistically useful sample of PMS stars of intermediate mass.

{A second motivation of this study is to explore the general role of magnetic fields in the late stages of formation of intermediate-mass stars. Accretion onto lower-mass PMS T Tauri stars is now generally supposed to be mediated by the presence of strong, large-scale magnetic fields (e.g. discussion by Johns-Krull et al. (1999) and references therein). Some authors (e.g. Vink et al. 2002, Muzzerole et al. 2004) have suggested that similar ``magnetospheric accretion'' may occur in intermediate-mass PMS stars as well. The observations presented here provide a unique opportunity to explore this proposal.}

Many intermediate-mass PMS stars are identifiable observationally as Herbig Ae/Be (HAeBe) stars (Herbig 1960; Cohen \& Kuhi 1979; Finkenzeller \& Mundt 1984; Hillebrand et al. 1992). HAeBe stars are characterised by spectral types
A and B with strong, often ubiquitous emission lines. They are distinguished
from the classical Ae/Be stars by their IR colours and frequent
presence within dust-obscured regions and association with nebulae
(Waters \& Waelkens 1998). According to standard stellar evolution theory, HAeBe stars should not
posses {deep} outer convection zones {which generate important quantities of outward-flowing mechanical energy.} Rather, these stars are expected to have convective cores surrounded
by { primarily} radiative sub-photospheric envelopes (Iben 1965; Gilliland
1986). However, since 1980, repeated observations of these stars
(e.g. Praderie et al. 1982; Felenbok et al. 1983; Catala et al. 1986;
Hamann \& Persson 1992) have shown that many are intensely active. In
particular, they display characteristics often associated with
magnetic activity and the presence of chromospheres or coronae (e.g. Skinner \& Yamauchi 1996; Swartz et al. 2005). These properties{ have been proposed as indicators that} at
least some of these stars { or their circumstellar envelopes} are intensely magnetically active. 

Previous attempts to detect magnetic fields in HAeBe stars have
produced varied results. Catala et al. (1993) attempted to detect
Zeeman circular polarisation in the Fe~{\sc ii} $\lambda$5018 and
He~{\sc i}~$\lambda 5876$ lines in the spectrum of the prototypical
HAeBe star AB Aur. High resolution, high signal-to-noise ratio spectra
of AB Aur resulted in no detection of a magnetic field, with upper
limits on the order of 1 kG. A survey undertaken by Glagolevski \&
Chountonov (1998), including a larger sample of stars, also found no
fields, but with relatively poor precision. On the other hand, Donati et al. (1997) and Donati
(2000) reported the probable and definite
(respectively) detection of a circular polarisation signature in
metallic lines of the Herbig Ae star HD 104237, providing the first
detection of a magnetic field in a HAeBe star. More recently, Hubrig et al. (2004, 2006a)
have claimed the detection of strong magnetic fields in several HAeBe stars. Clearly,
indirect and direct evidence for the presence of magnetic fields in some HAeBe stars in increasing, but it remains very sparse.

In this study, we use the low-resolution spectropolarimeter FORS1 at the ESO VLT to conduct a search for magnetic fields in about 50 HAeBe stars, the largest such sample ever studied. First results of this study have already been reported in a short paper by Wade et al. (2005).

\begin{figure}

\centering

\includegraphics[width=8cm]{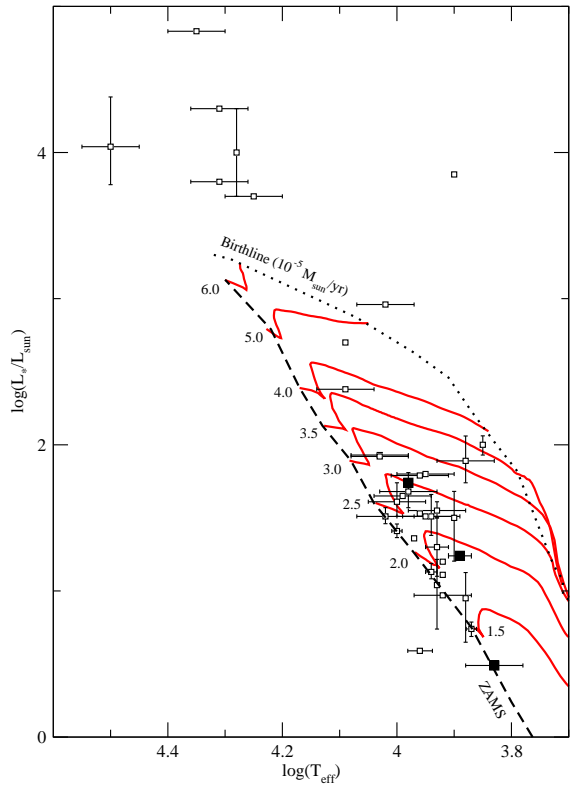}

\caption{HR diagram of sample stars with known $T_{\rm eff}$ and $\log L/L_{\odot}$, derived from Table~1. The data are interpreted using the model evolutionary calculations of Palla \& Stahler (1993), {which assume an accretion rate of $10^{-5}~M_\odot$/yr before the birthline.} Each evolutionary track is labelled with the corresponding stellar mass. Notably, a majority of the sample is concentrated between 1.5-3~$M_\odot$. { Stars in which magnetic fields are possibly detected in this study (HD 101412, BF Ori, and HD 36112) are indicated by filled symbols (the luminosity of CPD-53 295 is not known, and thus we are unable to place it on the HR diagram).}}\label{zBz}

\end{figure}

\section{Sample Selection\label{select}}

Our study required the selection of a relatively large number of HAeBe stars to allow us to derive statistically meaningful conclusions about the presence of magnetic fields in these stars, and the relationship of magnetic HAeBe stars to the Ap/Bp stars. Various literature sources were used for target selection, primarily the catalogues of HAeBe stars and HAeBe candidates by Th\'e et al. (1994) and by Vieira et al. (2003). 

The catalogue of Th\'e et al. (1994) contains six categories of stars; the stars selected for our study were obtained only from the first category, which contains stars historically known as HAeBe stars, or strong candidates of the group. According to the authors, all of these stars possess near- or far-infrared excess and emission lines, associated with the presence of circumstellar dust, { discs} and energetic ouflows which are usually found in HAeBe stellar environments. On the other hand, Vieira et al. (2003) produced a catalogue of HAeBe stars and strong candidates from an initial search for new T Tauri stars (pre-main sequence stars of lower mass) using the Infrared Astronomical Satellite (\textit{IRAS}) point source catalogue\footnote{\textit{http://irsa.ipac.caltech.edu/}}. Because the initial search was based on circumstellar dust properties, it included HAeBe stars along with the T Tauri stars. { Viera et al.} extracted the HAeBe stars by filtering the data using specific requirements such as a spectral type earlier than F5, emission at H$\alpha$, and a minimum level of infrared emission. The majority of the stars were associated by the authors with a star forming region (SFR). Based on the quality of these two literature sources and the arguments presented by their authors, we conclude that all of the stars in our sample are {\em bona fide} HAeBe stars. In total, 50 HAeBe stars have been selected, spanning in a spectral type range from F2 through B0.

The selected stars are listed in Table~1. The spectral types of the stars have been obtained from the bibliographic service \textit{SIMBAD} (unless indicated otherwise). Table~1 also contains physical properties (where available) such as mass, radius, effective temperature, luminosity, $v \sin i$, age and distance, found in the literature. Whenever the sources permitted, the uncertainties associated with these parameters were included. 

As can be seen in Table~1, { 7} objects have no well-determined physical parameters in the literature. In addition, the physical parameters (and in particular the luminosities) of many objects are highly uncertain. Notwithstanding this limitation, we have placed those stars with known temperatures and luminosities on the HR diagram in Fig. 1, along with the model pre-main sequence evolutionary tracks of Palla \& Stahler (1993). Based on Fig. 1, most of the stars in the observed sample have masses above 2~$M_\odot$, and fully 18 of the 37 stars on the diagram (49\%) have masses between 2 and 4~$M_\odot$. Eight stars have masses below 2~$M_\odot$, and 9 stars have masses larger than 4~$M_\odot$.

\setcounter{table}{0}

\begin{table*}\label{haebe}

\begin{center}

\begin{tabular}{lcccccccc}\hline\hline

Name & Sp. Type & Mass & $R_\ast$ &$\log (T_{\rm eff}$) & $\log (L_\ast/L_\odot)$ & $v \sin i$ & Age & Distance\\

& & ($M_\odot$) & ($R_\odot$) & (K) & & (km/s) & (Myr) & (pc) \\

\hline

CPD-53 $295^b$  & F2 & - & - & $^b$3.83 & - & $^b$175 & - & - \\
HD $17081^d$    & B7IV & $^*$4.6 & $^*$4.9 & $^o$4.09 & $^o$2.70 & $^n$18 & $^o$0.1  & $^o$135 \\
HD $278937^a$   & A3   & $^g$2.0 & $^*$1.9 & $^g$3.92 & $^g$1.21 & - & $^g$8.0 & $^g$350 \\
HD $275877^a$   & A2IIvar+& $^g$1.9 & $^*$1.7 & $^g$3.92 & $^g$1.12 & $^i$217 $\pm$ 13 & $^g$11.0 & $^g$120 \\
HD $31293^a$    & A0Vpe & $^f$2.4 & $^b$2.7 & $^f$3.98$\pm$0.05 & $^f$1.68$^{+0.13}_{-0.11}$ & $^f$80 & $^f$2.0 & $^b$160 \\
HD $31648^a$    & A3pshe+ & $^f$2.2 & $^*$2.5 & $^f$3.94$\pm$0.05 & $^f$1.51$^{+ 0.15}_{- 0.13}$ & $^i$102 $\pm$ 5 & $^f$2.5 & $^p$131 $\pm$ 20 \\
HD $293782^a$   & A3e & $^c$3.3 & $^c$3.2 & $^c$3.96$\pm$0.05 & $^c$1.79 & $^i$215 $\pm$ 15 & $^g$4.0 & $^c$460 \\
HD $34282^a$    & A0e+sh & $^j$1.39 & $^q$0.6 & $^j$3.94$^{+ 0.02}_{-0.01}$ & $^j$1.13$^{+0.06}_{-0.05}$ & $^j$110 $\pm$ 10 & $^j$6.4 & $^j$348 \\
HD 35187$^b$    & A2e & $^r$2.2 & $^*$2.3 & $^r$3.96 & $^r$1.53 & $^h$105 $\pm$ 9 & $^r$5.0 & $^r$150\\
HD $287841^a$   & A5III:e & $^m$1.55 $\pm$ 0.15 & $^*$1.4 & $^m$3.87 $\pm$ 0.01 & $^m$0.74 $\pm$ 0.05 & -  & $^*$20.0& - \\
HD $35929^a$    & A5 & $^s$3.6 $\pm$ 0.2 & $^*$6.6 & $^s$3.85 $\pm$ 0.01 & $^s$2.00 $^{+0.06}_{-0.07}$ & $^f$150 $\pm$ 30 & $^f$3.2 & $^s$390 $\pm$ 30 \\
HD $36112^a$   & A8e & $^{cc}1.8\pm 0.2$ & $^*$2.3 & $^q$3.89$\pm$0.02 & $^*$1.24& $^cc60\pm 6$ & $^q$6.0 & $^*$200 $^{+63}_{-40}$ \\
HD $244604^a$  & A3 & - & - & - & - & - & - & - \\
HD $245185^a$  & A5 & $^g$2.2 & $^*$1.8 & $^g$3.97 & $^g$1.36 & - & $^g$8.0 & $^g$400 \\
T Ori$^a$      & A3V &$^c$ 3.6 & $^c$2.8 & $^c$4.03$\pm$0.05 & $^c$1.92 & $^c$100 & $^g$1.7 & $^c$460 \\
V380 Ori$^a$   & A0 & $^{aa}2.8\pm 0.3$ & $^{aa}2.7$ & $^c$4.03$\pm$0.05 & $^c$1.93 & - & $^*$1.4 & $^c$460 \\
HD $37258^a$   & A2V & $^h$3.0 & - & $^h$3.96$\pm$0.02 & $^t$0.59 & $^h$160 $\pm$13 & - & - \\
BF Ori$^a$     & A5II-IIIevar & $^c$1.4 &$^c$ 1.3 & $^c$3.83$\pm$0.05 & $^c$0.49 & $^i$37 $\pm$ 2 &  $^*$25.0 & $^c$460 \\
HD $37357^a$   & A0e & - & - & $^b$3.96 & - & - & $^u$$\leq$6.0 & $^u$480 \\
HD $37411^a$   & B9V & - & - & $^h$4.03 & - & - & - & $^v$510 \\
HD $37490^a$   & BIIIe & $^c$16.9 & $^c$11.4 & $^c$4.31$\pm$0.05 & $^c$4.30 & $^c$160 & - & $^c$360 \\
HD $37806^a$   & A0 & $^h$3.0 & $^*$2.4 & $^f$3.95$\pm$0.05 & $^f$$>$1.51 & $^f$120 $\pm$ 30 & $^*$2.6 & - \\
HD $250550^a$  & B7e & $^c$4.8 & $^c$3.5 & $^c$4.09$\pm$0.05 & $^c$2.38 & $^c$110 & $^g$1.4 & $^c$700 \\
HD $259431^a$  & B6pe & $^c$12.2 & $^c$6.4 & $^c$4.31$\pm$0.05 &$^c$ 3.80 & $^c$100 & $^g$0.1 & $^c$800 \\
HD $52721^a$   & B2Vne & $^c$24.1 & $^c$17.4 & $^c$4.35$\pm$0.05 & $^c$4.83 & $^c$400 & - & $^c$1150 \\
HD $53179^a$   & Bpe & $^w$16 & - & $^l$3.90 & $^l$3.85 & - & $^w$0.3 & $^l$1150 \\
HD $53367^a$   & B0IV:e & $^e$13$\pm$3 & $^c$17.4 & $^e$4.50$\pm$0.05 & $^e$4.04$^{+0.34}_{-0.26}$ & $^c$30 & - & $^e$250$^{+120}_{-60}$ \\
NX Pup$^a$     & A0 & $^c$3.0 & $^c$2.4 & $^c$3.99$\pm$0.05 & $^c$1.65 & $^f$120 $\pm$ 10 & $^*$2.3 & $^c$450 \\
HD $68695^b$   & A0V & - & - & - & - & - & - & - \\
HD $72106A^b$   & A0IV & $^{aa}2.4\pm 0.4$ & - & $^{aa}4.04$ & $^{aa}1.3$ & $^{aa}45$ &-& $^b$290\\
HD $72106B^b$   & A0IV & $^{aa}1.75\pm 0.25$ & - & $^{aa}3.90$ & $^{aa}1.0$ &$~10$ &-&$^b$290\\
HD $76534^a$   & B2Vne & $^c$11.4 & $^c$7.5 & $^c$4.25$\pm$0.05 & $^c$3.70 & $^c$110 & - & $^c$870 \\
HD $85567^a$   & B5Vne & - & - & $^x$4.28 & $^x$4.0 $\pm$ 0.3 & $^x$70 & - & $^x$1500 $\pm$ 500 \\
HD $87403^b$   & A1III & - & - & - & - & - & - & - \\
HD $94509^a$   & A0Ib & - & - & - & - & - & - & - \\
HD $95881^a$   & A1/A2III/IV & - & - & - & - &  $^y$150 & - & - \\
HD $96042^b$   & B1V:ne & - & - & - & - & $^b$132 & - & - \\
HD $97048^a$   & A0pshe & $^c$3.4 & $^c$2.5 & $^f$4.00$\pm$0.05 & $^f$1.61$^{+0.13}_{-0.10}$ & $^f$140 $\pm$ 20 & $^f$$>$6.3 & $^c$215 \\
HD $98922^a$   & B9Ve & $^*$5.0 & $^*$9.2 & $^f$4.02$\pm$0.05 &$^f$$>$2.96 & - & $^*$0.1 & - \\
HD $100453^b$  & A9Ve & $^k$1.7 & $^*$1.7 & $^k$3.86 & $^k$0.95$^{+0.18}_{-0.30}$ & - & $^*$10.0 & $^k$114 \\
HD $100546^a$  & B9Vne & $^e$2.4 & $^*$1.7 & $^e$4.02$\pm$0.05 & $^e$1.51$^{+0.06}_{-0.05}$ & - & $^e$$\sim$10 & $^e$103$^{+7}_{-6}$ \\
HD $101412^a$  & B9.5V &$^{aa}2.6\pm 0.3$  & - & $^{b}3.98$& 1.74 & $^{dd}$7 & 2 & $^{bb}500-700$ \\
HD $104237^a$  & A:pe & $^f$2.3 & $^*$2.7 & $^f$3.93$\pm$0.05 & $^f$1.55$^{+0.06}_{-0.05}$ & - & $^f$6.3& - \\
HD $132947^a$  & A0 & - & - & - & - & - & - & - \\ 
HD $141569^a$  & B9.5e & $^j$2.0 & $^*$1.7 & $^j$4.00$^{+ 0.01}_{- 0.01}$ & $^j$1.41$\pm$0.05 & $^j$236 $\pm$ 15 & $^j$4.71 & $^j$108 \\
HD $142666^a$  & A8Ve & $^k$1.8 & $^*$1.5 & $^k$3.93 & $^k$1.04$^{+ 0.18}_{- 0.30}$ & $^i$72 $\pm$ 2 & $^*$10.0 & $^k$116 \\
HD $144432^a$  & A9/F0V & $^k$2.2 & $^*$3.0 & $^k$3.90 & $^k$1.51$^{+ 0.18}_{- 0.30}$ & $^i$85 $\pm$ 4 & $^*$2.0 & $^k$200 \\ 
HD $144668^a$  & A7IVe & $^e$3.1 $\pm$ 0.5 & $^c$3.9 & $^e$3.88$\pm$0.05 & $^e$1.89$^{+0.17}_{-0.15}$ & $^h$180$\pm$50 & $^e$0.6$\pm$0.4 & $^e$210$^{+50}_{-30}$ \\
Ty Cra$^a$     & B8e & $^c$1.9 & $^c$1.5 & $^c$3.92$\pm$0.05 & $^c$0.97 & $^c$88 & $^*$11.5 & $^c$130 \\
HD $190073^a$  & A2IVpe & $^*$2.7 & $^*$3.3 & $^f$3.95$\pm$0.05 & $^f$$>$1.80 & $^z$20 & $^*$
1.3 & $^f$$>$290\\
\hline

\end{tabular}

\end{center}

\caption[]{HAeBe programme stars { ordered by right ascension}, including spectral type, mass, radius, effective temeperature, luminosity, $v\sin i$, age and heliocentric distance, where available. Reference: $^a$Th\'e et al. (1994), $^b$Vieira et al. (2003), $^c$Hillenbrand et al. (1992), $^d$Malfait et al. (1998), $^e$van den Ancker et al. (1997), $^f$van den Ancker et al. (1998), $^g$Hern\'andez et al. (2004), $^h$B$\ddot{\rm{o}}$hm \& Catala (1995), $^i$Mora et al. (2001), $^j$Mer$\rm{\acute{\i}}$ et al. (2004), $^k$Dominik et al. (2003), $^l$Fuente et al. (2002), $^m$Pinheiro et al. (2003), $^n$Acke \& Waelkens (2004), $^o$Habart et al. (2003), $^p$Galazutdinov et al. (2003),$^q$Mannings \& Sargent (2000), $^r$Natta et al. (2004), $^s$Marconi et al. (2000), $^t$Kovalchuk et al. (1997), $^u$Holmes et al. (2003), $^v$Hovhannessian et al. (2001), $^w$van den Ancker et al. (2004), $^x$Miroshnichenko et al. (2001), $^y$Grady et al. (1996), $^z$Carporon \& Lagrange (1999), $^{aa}$Wade et al. (2005), $^{bb}$Corradi et al.  (1997), $^{cc}$Beskrovnaya et al. (1999), $^{dd}$Guimaraes et al. (2006), $^*$This work.}

\end{table*}

\begin{figure*}[ht]

\centering

\includegraphics[width=9cm]{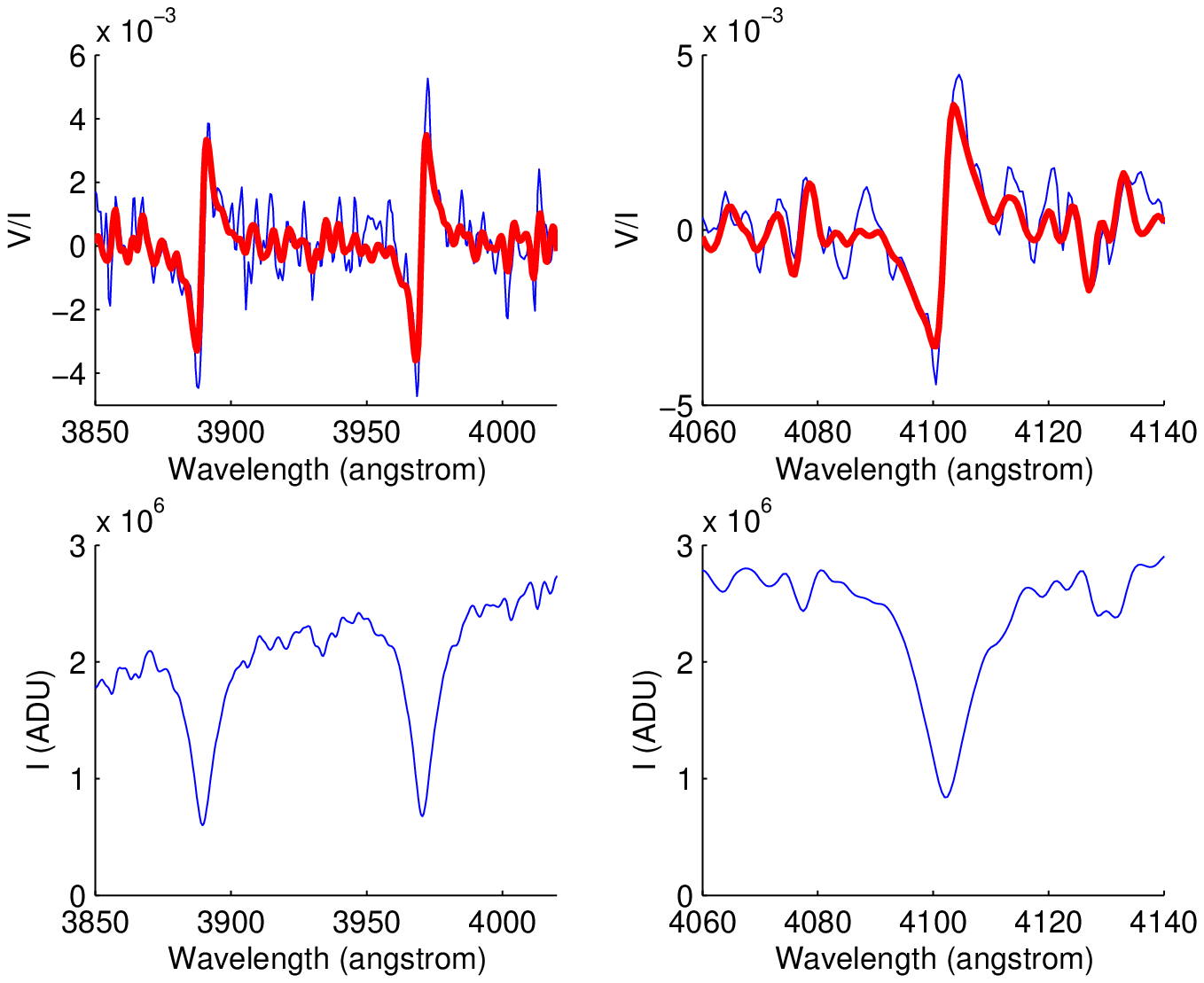}\includegraphics[width=9cm]{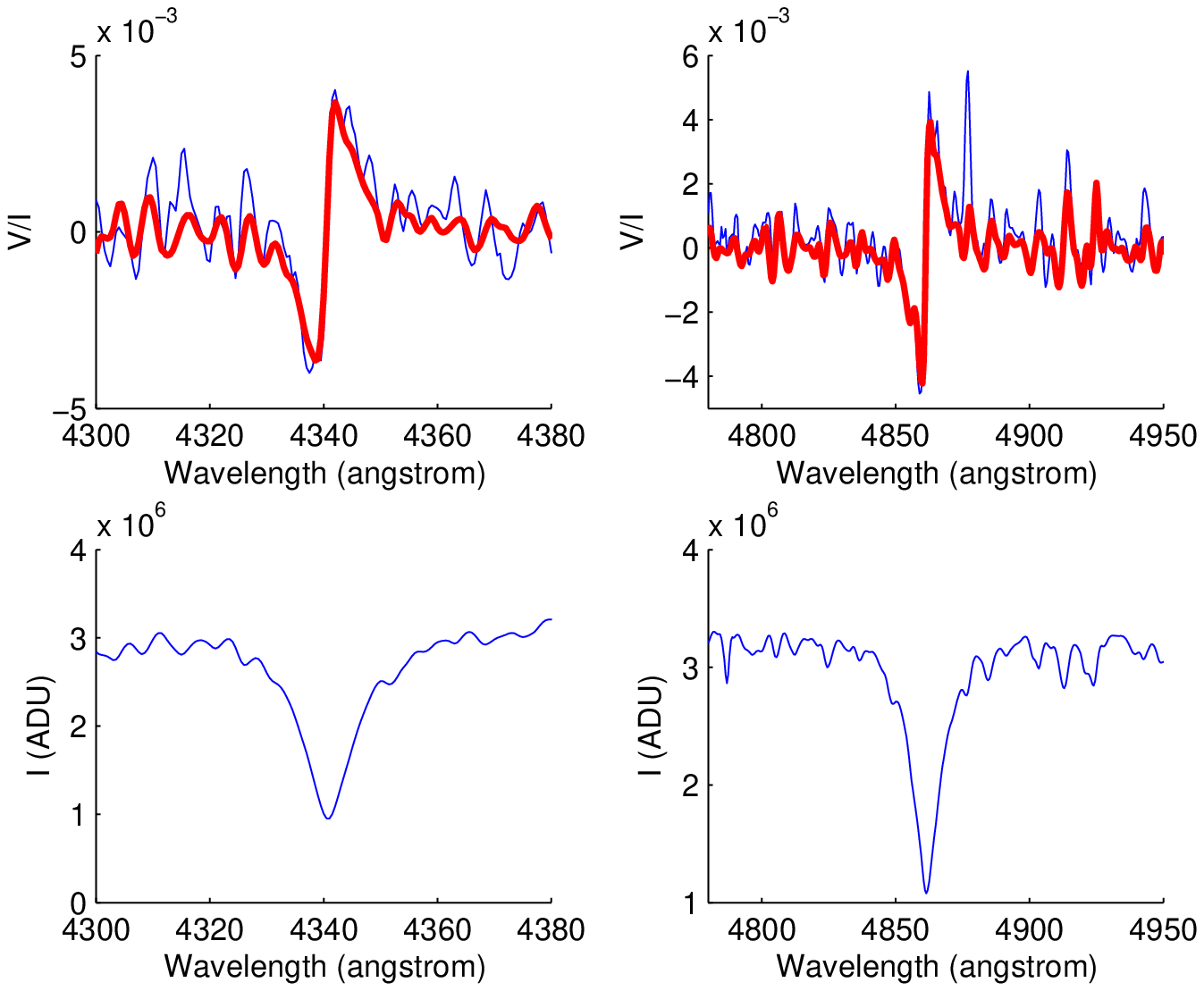}

\caption{Stokes $I$ and $V$ profiles in the FORS1 spectrum of HD 94660. From left to right, individual frames show H8, Ca~{\sc ii} H and K and H$\epsilon$; H$\delta$; H$\gamma$; and H$\beta$. Thin lines represent the observed spectrum, while thick lines represent the computed Stokes $V$ spectrum determined using Eq. (1). Note the strong, clear signatures associated with all Balmer lines.}

\end{figure*}

\section{Observations and data reduction}

The sample stars were observed at the European Southern Observatory (ESO) Very Large Telescope (VLT) with the first of the two FOcal Reducer and low dispersion Spectrographs (FORS1). FORS1 was mounted at the Cassegrain focus of one of the telescope units and was used in spectropolarimetric {\em fast} mode. A total of about five nights were allocated for our study, distributed over three observing runs (visitor mode runs in February (P72) and November 2004 (P74), and one {service mode run of} Director's Discretionary Time in April 2004). 

{The large majority of our spectra were obtained with the grism 600 B, which covers the spectral range from 3450 to 5900\,\AA, and yields a dispersion of 1.2\,\AA\ per pixel.
Spectra were obtained with a slit width of 0.5" or 0.8", yielding a resolving power of 1560 or 975, respectively. In a few cases, grism 1200\,g (now decommissioned) was used, which covers the spectral range from 4310 to 5490~\AA\ (which includes H$\beta$ and H$\gamma$ only). Its dispersion is 0.6\,\AA\ per pixel, and with a slit width of 0.5" or 0.8", this grism yields a resolving power of 2800 or 1800, respectively.}

Sixty-eight observations of the 50 targets were obtained in circularly polarised light following the procedure {described} by Bagnulo et al. (2006). The light gathered by the telescope is sent through the {Longitudinal Atmospheric Dispersion Corrector and the instrument collimator, then passes through} a single slit into the top section of the FORS1 instrument. There, the stellar light passes through a rotatable quarter-wave plate (located in the filter/camera section of FORS1) oriented at $\pm 45\degr$ with respect to the fast axis of the analyser, which converts the circularly polarised light into linearly polarized light. The orthogonal linearly polarized beams are then separated by the Wollaston prism before being dispersed by a grism. Two spectra (each corresponding to one of the oppositely polarised beams) are finally recorded simultaneously on the CCD. By changing the orientation of the quarter-wave plate with respect to the Wollaston fast axis (alternately $+45\degr$ or $-45\degr$) and obtaining an additional exposure, the positions of the { right and left circularly polarised} beams are exchanged throughout the instrument, and in particular their respective positions on the CCD are interchanged. As described {in the FORS1 manual (VLT-MAN-ESO-13100-1543)}, this procedure allows for the removal of systematic errors due to CCD inhomogeneities and differential wavelength calibration errors.
  
In order to obtain a high signal-to-noise ratio, multiple exposures of each star were obtained. An even number of exposures (between 2 and 30, depending on the apparent magnitude of the star and the seeing) was obtained, each pair of exposures corresponding to a quarter-wave plate orientation of either $+45^{\rm{o}}$ or $-45^{\rm{o}}$. The data were reduced using the procedure described in detail by Bagnulo et al. (2006).

The journal of observations (Table~2) reports our observations of both standard and programme stars. It should be noted that grism 600B was used for all observations unless indicated otherwise (in which case grism 1200g was used). Table~2 includes the Julian Date of mid-observation, the slit width, exposure time and number of exposures, and the resultant { peak} signal-to-noise ratio { per \AA}. The table also contains the results of the magnetic field diagnosis, which are discussed in Sect. 4.

\section{Results}

The longitudinal magnetic field has been inferred from the polarisation spectra in the manner described by Bagnulo et al. (2006). Briefly, the field diagnosis is obtained using a Least-Squares fit based on the predicted circularly polarised flux $V/I$ in the weak-field regime, given by

\begin{equation}
\frac{V}{I} = - g_\mathrm{eff} \ \cz \ \lambda^{2} \
                \frac{1}{I} \
                \frac{{\rm d}I}{{\rm d}\lambda} \
                \Bz\;,
\label{EqBz}
\end{equation}

\noindent where $g_\mathrm{eff}$ is the effective Land\'{e} factor, \textit{I} is the usual (unpolarised) intensity, $\lambda$ is the
wavelength expressed in \AA, \Bz\ is the longitudinal field
expressed in gauss, and
\[
\cz = \frac{e}{4 \pi m_\mathrm{e} c^2}
\ \ \ \ \ (\simeq 4.67 \times 10^{-13}\,\mathrm{\AA}^{-1}\mathrm{G}^{-1})
\]

\noindent where $e$ is the electron charge, $m_\mathrm{e}$ the electron mass, and $c$ the
speed of light. For the atmosphere of a main sequence A or B type star, the weak-field
approximation holds, for H lines, up to around 10-20\,kG, and, for metal lines,
up to about 1\,kG.

To derive the longitudinal field, we minimise the expression
\begin{equation}
\chi^2 = \sum_i \frac{(y_i - \Bz\,x_i - b)^2}{\sigma^2_i}
\label{EqChiSquare}
\end{equation}
where, for each pixel $i$ in the reduced spectrum, $y_i = (V/I)_i$, $x_i =
-g_\mathrm{eff} \cz \lambda^2_i (1/I\ \times \mathrm{d}I/\mathrm{d}\lambda)_i$, and $b$
is a constant term. Uncertainties $\sigma_i$ associated with each reduced spectral pixel are
obtained as described by Bagnulo et al. (2006). The longitudinal field uncertainty { is obtained from the formal uncertainly of the linear regression, and its derivation is described by Bagnulo et al. (2002).}

{As is discussed later in this Section, we encountered a number of cases for which the field was detected at
about the 3-$\sigma$ level, and in which minor changes in the data
reduction would transform a marginal detection in a null or into a
definite detection. Although these cases should certainly be
investigated further via additional observations, we tried to extract further
information from the available spectra, to formulate a more robust and
reliable criterion for field detection.

Each observation block consisted of several pairs of observations
obtained with the retarder waveplate at $-45\degr$ and $+45\degr$. As
explained by Bagnulo et al. (2006), there are two ways to determine
\Bz.\ {\em(i)} One can combine all observations to obtain a final,
high signal to noise ratio Stokes $V$ profile, from which is determined
a unique value of \Bz. {\em(ii)} Alternatively, one can combine individual pairs
of frames obtained at $-45\degr$ and $+45\degr$ to obtain several
lower signal-to-noise ratio Stokes $V$ profiles, which yield several $\Bz_i$
determinations. The final \Bz\ value is computed from the average of these
$\Bz_i$ values. If both methods yield a detection at the $3\,\sigma$ level or
higher, we flag the observation with "D" (implying a definite detection according to both criteri). 
If a $3\,\sigma$ detection is obtained with only one of the two methods, we flag the observation
with "d". Finally, if both methods yields a null detection ($< 3\,\sigma$) we
flag the observation with "n".

This procedure has been repeated four different ways:
using the whole spectrum, using only metal (and He) lines, using all Balmer
lines, and using all Balmer lines except H$\beta$. (In our grism 600B observations, H$\beta$ typically exhibits the strongest emission of any of the Balmer lines, with emission intensity decreasing strongly for higher series members. Measurement of the field using only Balmer lines H$\gamma$ and higher is an attempt to mitigate the effects of this emission.) A flag "n", "d", or
"D", as defined above, was assigned to each kind of analysis. It is
clear that a series "DDDD" denotes a certain detection, a "nnnn"
series denotes a null detecion, whereas intermediate cases (including
possibly "d" flags) indicate the need of further observations to fully
asses whether the star is magnetic or not. The results of the four diagnoses of the magnetic field are summarised in columns 6-9 of Table 2, with their significance summarised in column 10.}

\subsection{Standard stars}

In order to evaluate the nominal operation of the instrument, and the accuracy of the reduction and analysis techniques, we observed both magnetic and non-magnetic standard stars. The magnetic Ap star HD 94660 (spectral type A0p) was observed 4 times: {twice (consecutively)} during the P72 run with grism 600B, and twice during the P74 run with both grism 600B and grism 1200g. The non-magnetic A1V star HD 96441 was observed during the DDT run. The magnetic diagnosis for these standard stars is included in Table 2.

\begin{figure*}

\centering

\includegraphics[width=18cm]{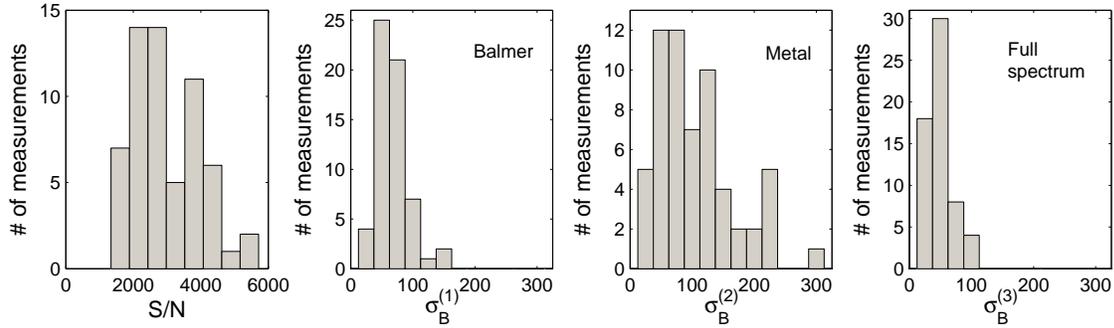}
\caption{Histograms of the signal-to-noise ratio (S/N, column 5 of Table 2) and formal uncertainties of the longitudinal field measurements (columns 6-8 of Table 2).}\label{zBz}
\end{figure*}

\begin{figure*}

\centering

\includegraphics[width=15cm]{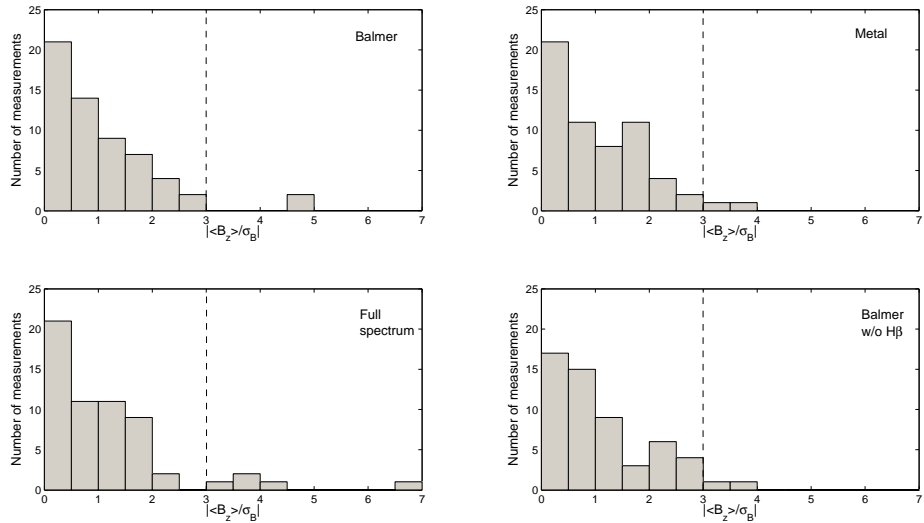}

\caption{Histograms of the detection significance $z=\langle B_z\rangle/\sigma_B$ of magnetic field measurements of programme stars. Each panel corresponds to one of the 4 measurements reported in columns 6-9 of Table 2. The dashed line indicates the $3\sigma$ detection threshold.}\label{zBz}

\end{figure*}
\begin{figure*}

\centering

\includegraphics[width=9cm]{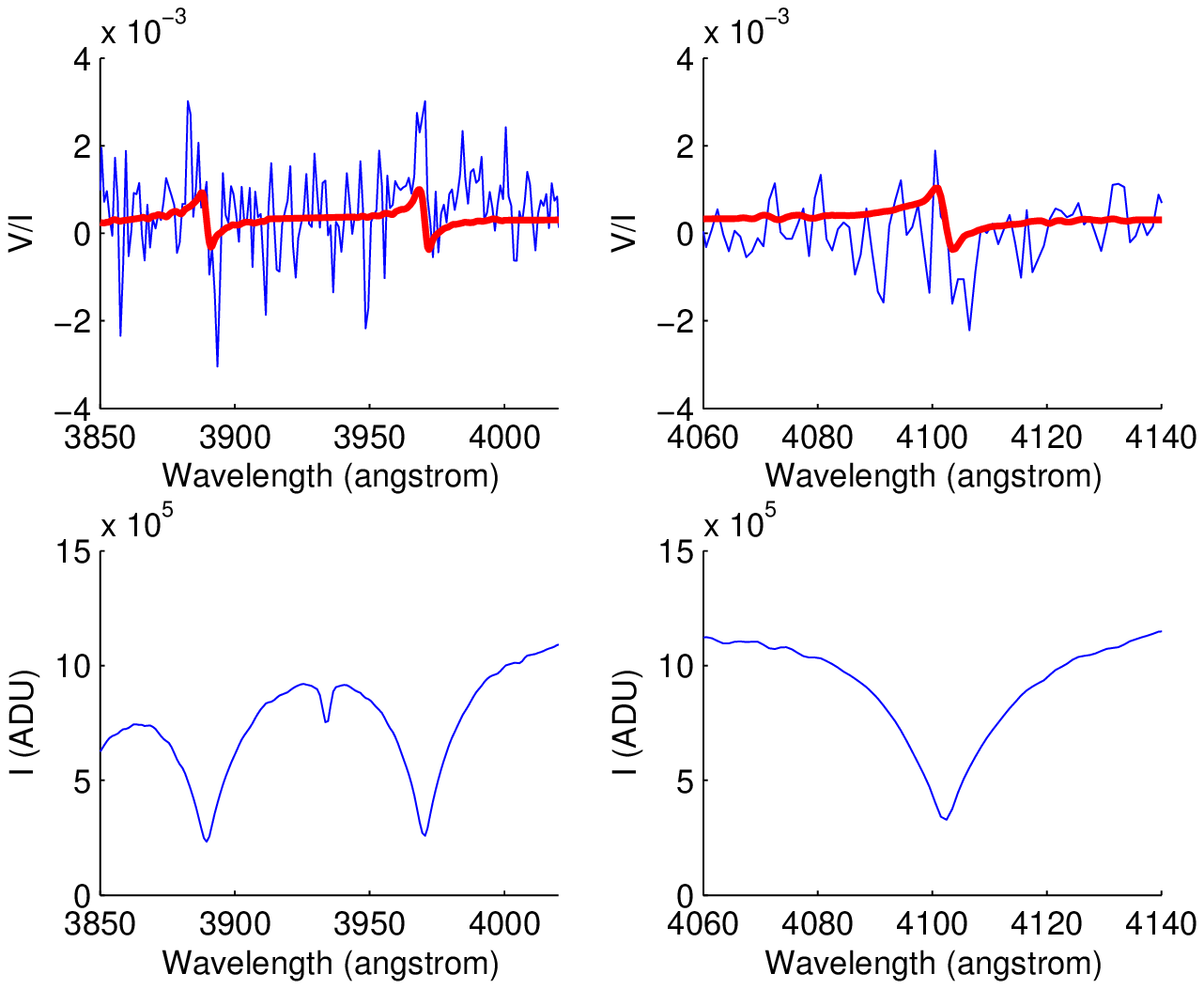}\includegraphics[width=9cm]{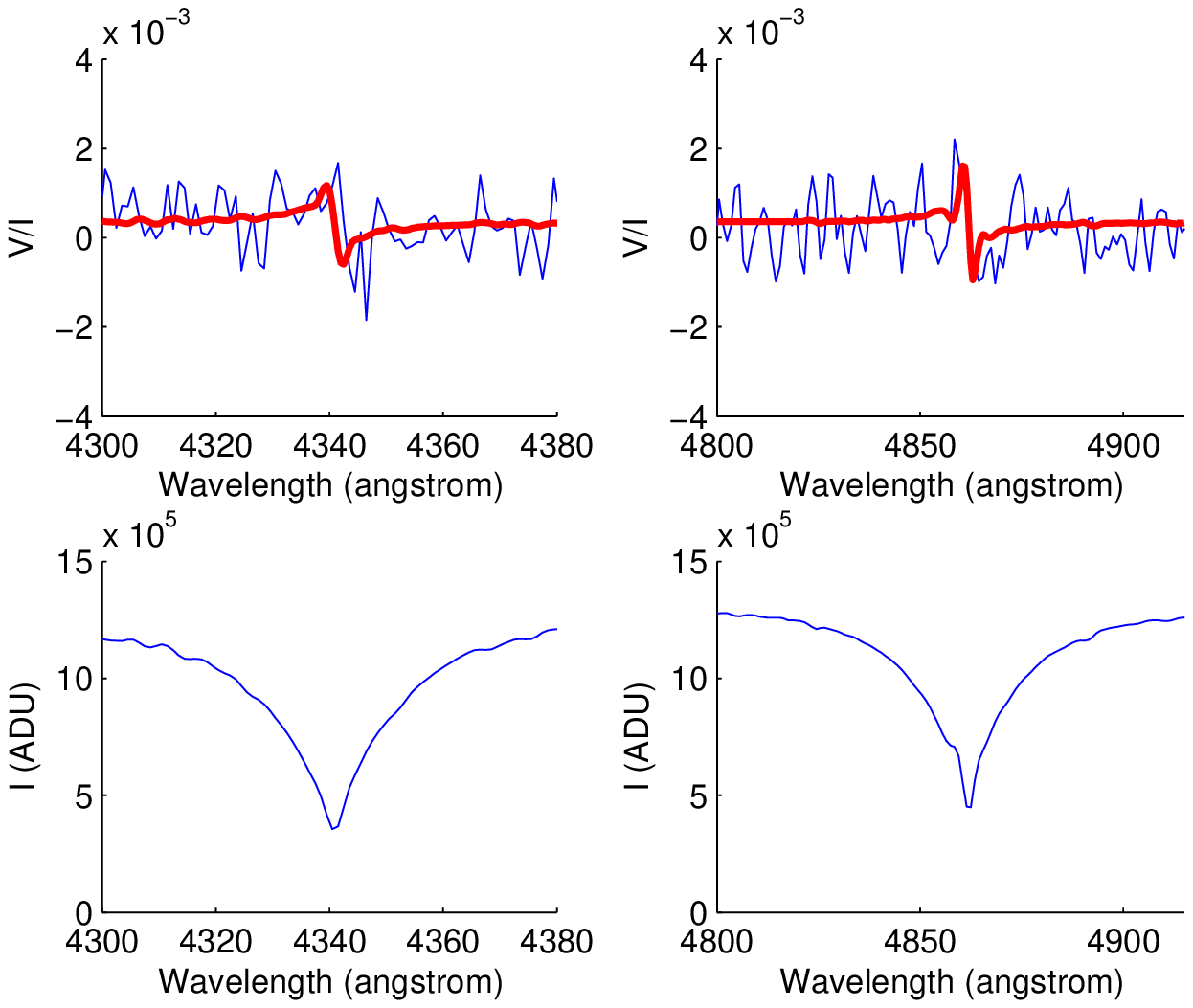}
\caption{Stokes $I$ and $V$ profiles in the spectrum of HD 101412. }

\end{figure*}

In P72, the field of HD 94660 measured in Balmer lines was { $-2534\pm 63$~G} and {$-2429\pm 70$~G}, and the fields measured in all Balmer lines except H$\beta$ were $-2363\pm 78$~G { and $-2535\pm 71$~G}. The consistency of these results is expected for a main sequence Ap star with no Balmer line emission. On the other hand, the fields measured in metal lines were significantly weaker, { $-1617\pm 94$~G and} $-2026\pm 54$~G. As discussed by Bagnulo et al. (2006), inconsistency between metal line and Balmer line longitudinal fields is common in stars with strong fields, and likely results from a breakdown of the weak-field assumption (in the case of metallic lines), possibly coupled with the influence of nonuniform surface distributions of metal abundances. Moreover, these two measurements were obtained using different slit widths (0.5 vs. 1 arcsec). To explore the possible effects of the slit size difference, we have performed a few numerical experiments, computing polarised synthetic spectra and degrading the resolution of both Stokes $I$ and $V$ to an adjustable value. We computed the binned and smoothed $V$ and $dI/d\lambda$, and calculated the slope and inferred longitudinal field from the regression. The interesting result is that as the resolving power degrades below about 2000, the deduced field from metallic lines starts to change a lot. It generally gets smaller, but it varies in a somewhat unpredictable manner. This appears to be basically because, as we reduce the resolution, we are going from the regime where each depression in the observed spectrum is roughly one line to the regime where each depression is one blend of nearby strong lines, and treating this as a single line is an increasingly bad approximation. On the other hand, changing the resolving power will have negligible effect on Balmer lines (as observed). From this we conclude that it is best to use the highest resolving power possible (i.e. a 0.5 arcsec slit) to get the most out of the metallic line spectrum, and that probably the difference between the two metallic line measurements of HD 94660 is due to some combination of the seeing and the slit size. 

In P74, the field measured in Balmer lines from both the grism 600B and grism 1200g observations are consistent, and equal to $-2437\pm 41$~G (600B) and $-2337\pm 40$~G (1200g). The fields measured in metal lines ($-2074\pm 44$~G and $-2144\pm 16$~G) are also in reasonably good agreement.

The results for HD 94660 are fully consistent with those described by Bagnulo et al. (2006). In particular, the fields measured during P74 agree to within $1\sigma$ with the fields reported by them. Stokes $I$ and $V$ profiles of HD 94660 are illustrated in Fig. 2.

The single observation of HD 96441, obtained during the DDT run, was consistent with zero field ($-108\pm 43$~G in Balmer lines, and $78\pm 65$~G in metal lines).

\subsection{Programme stars}

\subsubsection{Data quality}

Two hundred and seventy-two magnetic field measurements have been obtained from the 68 observations of the 50 programme stars. The mean { peak} S/N { (per angstrom)} of the spectra was 2990:1, but varied significantly from 1340:1-5720:1 due to seeing and transparency conditions, and apparent brightness of the target. The range of S/N results in a relatively large range of longitudinal field uncertainties (all uncertainties in this paper are at 1$\sigma$ confidence). For measurements obtained from Balmer lines only, the mean uncertainty was 66~G, ranging from 19 to 162~G. For metal lines only, the mean uncertainty was 105~G, ranging from 21 to 291~G. For the full spectrum, the mean uncertainty was 48~G, ranging from 16 to 102~G. These distributions are illustrated in Fig. 3 (the distribution for all Balmer lines excluding H$\beta$ was very similar to that of all Balmer lines, and so is not illustrated here).

The distribution of error bars for the metal line measurements is somewhat broader than for the Balmer line or full spectrum regressions. As pointed out by Bagnulo et al. (2006), this is due to the fact that Balmer lines have similar strength in most of these stars, whereas metal lines change greatly from star to star. Although it is true that the diversity of Balmer line profiles in spectra of HAeBe stars is somewhat greater than for main sequence A and B stars, the responsible emission effects are mostly visible in H$\alpha$, and to a lesser degree in H$\beta$. At the resolution of FORS1, the profiles of higher members of the series are very similar to those of main sequence stars.

The magnetic field diagnoses obtained for the programme stars are summarised in Fig. 4, where we show histograms of the detection significance $z=|\langle B_z\rangle|/\sigma_B$. Each frame in Fig. 4 corresponds to one of the 4 measurements reported in Table 2. {Only 12} of the 272 programme star measurements (corresponding to {only 4 stars}) indicate a magnetic field detection significant at greater than or equal to the $3\sigma$ level. These stars are BF Ori (which contributes detections between 3.2$\sigma$ and 6.9$\sigma$ to all 4 frames, ``dddd''), HD 101412 (which contributes detections from 4.0$\sigma$ to 4.6$\sigma$ to 3 frames, ``DddD''), CPD-53 295 (which contributes detections of 4.0$\sigma$ to 2 frames, ``nDDn''), and HD 36112 (which contributes detections from 3.0$\sigma$-3.6$\sigma$ to 3 frames, ``nddn''). According to the flag in column 10 of Table 2, only HD 101412 { and CPD-53 295} indicate a significant field detected according to both criteria of Bagnulo et al. (2006) (and are therefore denoted with the symbol ``D''). No other observation of any programme star resulted in any field detected above the 3$\sigma$ level.

\subsubsection{Detected targets}

\vspace{5mm}
{\noindent \bf HD~101412}
\vspace{5mm}

HD 101412 (He3-692, PDS 057) was selected from the work of Th\'e et al. (1994) where it is noted as an historically known or strong candidate HAeBe star. It was first classified as a HAeBe star by Hu and Zhou (1990) based on its IRAS infrared excess. It exhibits double-peaked H$\alpha$ emission, as well as O~[{\sc i}]~$\lambda\lambda$6300, 6364 emission (e.g. Vieira et al. 2003), and may exhibit polarisation variability on timescales of minutes (Yudin \& Evans 1998). Although HD 101412 is not associated with a reflection nebula (Ray \& Eisloeffel 1994), Vieira et al. (2003) were able to associated it with the DC 295.0+1.3 star forming region (SFR) in the Centaurus and Crux complex. It therefore appears that HD 101412 is a bona fide HAeBe star and therefore a pre-main sequence star of intermediate mass. 

The surface temperature of HD 101412 reported by Vieira et al. (2003) is about 9500 K, too cool for HD 101412 to be a misclassified Be star. No parallax data is available for this star, although based on the reported distance to the DC 295.0+1.3 SFR (500-700 pc; Corradi et al. 1997), Wade et al. (2005) estimated its position on the HR diagram. This position, interpreted using the model evolutionary tracks of Palla \& Stahler (1993), indicates that HD 101412 has a mass of $2.6\pm 0.3~M_\odot$, and appeared at the birthline only about 2 Myr ago. This suggests that this star is still accreting material and burning deuterium as its main source of energy, and its structure should be significantly different than that of main sequence A stars.

HD 101412 was observed during the P72 observing run. Unfortunately, {observations of this star were obtained while clouds were passing}, and only 4 exposures were obtained, {resulting in some frames with low S/N and others that were partially saturated. The reduced spectrum (which has lower (1350:1) S/N, and is illustrated in Fig. 5)} shows strong Balmer lines which in general show little evidence of emission. H$\beta$ shows an asymmetry in the inner blue wing which is presumably due to emission. The Balmer lines are much broader than those of HD~94660 { (which has $T_{\rm eff}=10750$~K according to $uvby\beta$ photometry)}, supporting the cooler temperature derived by Vieira et al. (2003). If we assume this temperature, the wings of the H$\gamma$ and H$\delta$ profiles are well-fit using ATLAS9 Balmer profiles computed for $\log g=4.0$. The metallic-line spectrum is substantially weaker than that of HD~94660, although the Ca~{\sc ii} K line (first panel of {Fig. 5}) is relatively strong. The Mg~{\sc ii}~$\lambda 4481$ line and Fe~{\sc ii} lines of multiplet 42 ($\lambda\lambda\lambda 4923, 5018, 5169$) are also prominent. 

The longitudinal magnetic field has been diagnosed using both Balmer lines and metallic lines (flag ``DddD''). According to both of the detection criteria of Bagnulo et al. (2006), a field is detected in the Balmer lines: at the 4.6$\sigma$ level ($512\pm 111$~G) from the Balmer line regression based on the final reduced Stokes $I$ and $V$ spectra, and the $3.2\sigma$ level based on the weighted average of regression results obtained from the 4 pairs of Stokes $I/V$ spectra ($572\pm 181$). In addition, each of the 4 pairs of spectra produce Balmer line regression results that are consistent to within 1$\sigma$ with the result reported in Table 2. Although no clear Stokes $V$ signatures are evident in Balmer lines shown in {Fig. 5}, there does appear to be a reasonably convincing tendancy for the circular polarisation to be { positive} in the blue wing { and negative} in the red wing of the line (consistent with a positive longitudinal magnetic field)\footnote{{As explained at the end of Sect.~8 and in Fig.~8 of Bagnulo et al. (2006), in their survey of cluster magnetic Ap stars they encountered
numerous cases where a simple inspection of the $V/I$ profiles did not show
any obvious polarization signal, but in which the magnetic field was detected by
the linear correlation between $V/I$ and the quantity $1/I\
\times {\rm d}I/{\rm d} \lambda$. The reliability of this result is supported
by the histograms of Fig.~7 of Bagnulo et al. (2006), showing that virtually
no detections are found in "normal" A stars, where we do not expect to find
any magnetic field. Magnetic fields were detected only in Ap stars, i.e.,
where they are expected to be found.}}.

The full spectrum regression also indicates a longitudinal field detection, but only according to one of the criteria. On the other hand, the regression using only metallic lines, although corresponding to a single-criterion detection, shows similar magnitude but opposite sign. Although this could be attributed to the larger error bar resulting from the weak and noisy metallic-line spectrum, and differs from the Balmer-line result by only 2.3$\sigma$, it motivates us to be more cautious about claiming the existence of a field in this star. For the moment, we conclude tentatively that HD~101412 hosts a photospheric magnetic field with a longitudinal intensity of order 500~G, but further observations are clearly necessary.

\begin{figure*}

\centering

\includegraphics[width=9cm]{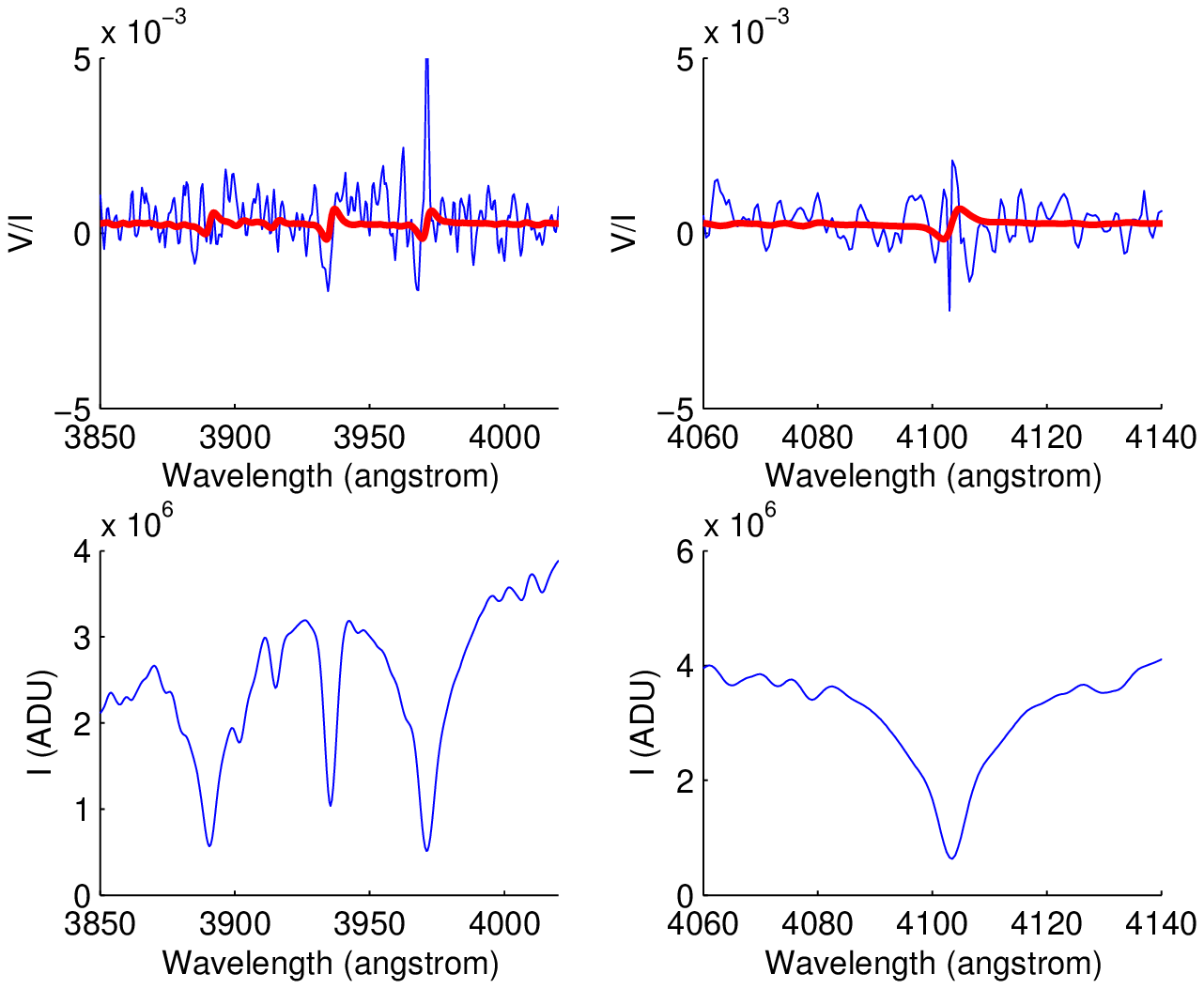}\includegraphics[width=9cm]{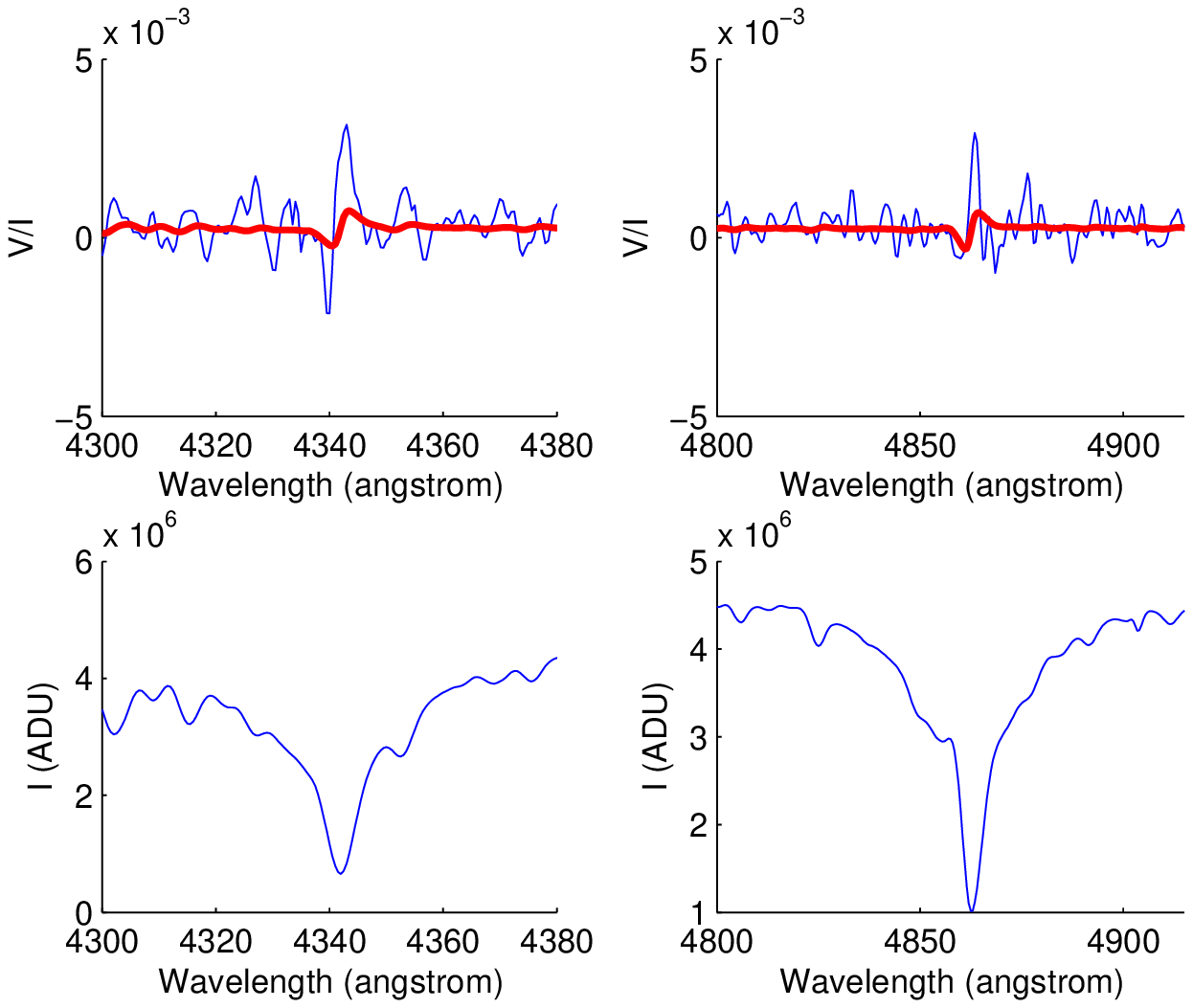}

\caption{Stokes $I$ and $V$ profiles in the spectrum of BF Ori. }

\end{figure*}

\vspace{5mm}
{\noindent \bf CPD-53 295}
\vspace{5mm}

{
CPD-53 295 (PDS 002) was selected from the work of Vieira et al. (2003), who were unable to associate it with a specific star-forming region. SIMBAD indicates a spectral type of F2, and Vieira et al. report an effective temperature of 6750~K and a rotational velocity of 175~\kms. Otherwise, this star is very poorly studied, appearing in only 5 articles in the ADS since 1970.  

CPD-53 295 was observed on a single night during the P74 observing run with both grisms, obtaining S/Ns of 2925:1 (grism 600B) and 1800:1 (grism 1200g). The spectra are rich in metallic lines, show Balmer lines in absorption, and very strong Ca~{\sc ii} H and K lines, consistent with the late spectral type. Neither the grism 600 nor the 1200g spectrum Stokes $V$ spectrum shows any signatures.

The longitudinal magnetic field has been diagnosed using both Balmer lines and metallic lines (flags ``nnnn'' and ``nDDn'', {the latter with grism 1200g}). According to both of the detection criteria of Bagnulo et al. (2006), no field at all is detected in the 600B observation. On the other hand, a field is detected in the metal lines and full spectrum in the 1200g observation, both at the $4.0\sigma$ level ($139\pm 35$~G, $129\pm 32$~G). When this spectrum is examined by eye, no significant signatures are evident.

Although we are unable to simply discard the apparent metallic-line detection in the 1200g spectrum, given that both observations of this star were obtained within about 1 hour of each other, and that the Balmer line diagnosis of the 1200g observation is inconsistent with the metallic line diagnosis, we do not consider the detection of CPD-53 295 to be convincing evidence of the presence of a magnetic field.

\vspace{5mm}
{\noindent \bf BF ORI}
\vspace{5mm}

BF Ori (BD-06$\degr$1259) was selected from the catalogue of Th\'e et al. (1994). This object is a well-known and extensively studied pre-main sequence star, exhibiting large infrared excess, H$\alpha$ emission, and polarimetric variability. Hillenbrand et al. (1992) derive an effective temperature of 6750~K, and a mass of 1.4~$M_\odot$, substantially lower than indicated by the SIMBAD spectral type (A5 II-IIIevar) and derived by Grinin et al. (2001: 8750~K). The challenge of classifying this star results, at least in part, from the presence of complex non-photospheric absorption lines (Hernandez et al. 2004). BF Ori is within 1$\arcmin$\ of an X-ray source identified by Damiani et al. (1994). According to Boehm \& Catala (1994), this star does not exhibit O~[{\sc i}]~$\lambda\lambda$6300, 6364 emission.

BF Ori was observed during the P72 observing run. Eight exposures were obtained, yielding a S/N of 2900:1. The spectrum (illustrated in Fig. 6) is characterised by strong, asymmetric Balmer lines (a systematic asymmetry can be followed down to at least H10, and possibly H11), and a rich, strong metallic-line spectrum. The Fe~{\sc ii} lines of multiplet 42 are especially prominent, and appear to show weak emission in the wings. The Ca~{\sc ii}~K line is very strong. Although it is at the very edge of our spectral domain, we see quite clearly an absorption profile of He~{\sc i}~$\lambda 5876$\footnote{{The detection of this strong absorption is confirmed from examination of high-resolution spectra in our possession. The detection of H~{\sc i}~$\lambda 5876$ is remarkable given the assumed temperature of this star, and it will be explored in detail in a future paper.}}.

The magnetic diagnosis for this star is summarised by the flag ``dddd'', indicating a detection according to one of the criteria for each of the four measurements ($-180\pm 38$~ (4.7$\sigma$) from Balmer lines, $-95\pm 30$~G (3.2$\sigma$) from metal lines, $-144\pm 21$~G (6.9$\sigma$) from the full spectrum, and $-158\pm 45$~G (3.5$\sigma$) from all Balmer lines excluding H$\beta$. In Fig. 6, noisy but apprently significant circular polarisation is present in all Balmer lines. Based on the detection of consistent longitudinal magnetic fields in both Balmer lines and metal lines, we tentatively conclude that BF~Ori hosts a magnetic field with a longitudinal intensity of order 150~G. However, as in the case of HD~101412, further observations are clearly necessarily in order to unambiguously establish the presence of a magnetic field.


\vspace{5mm}
{\noindent \bf HD 36112}
\vspace{5mm}

\begin{figure*}

\centering

\includegraphics[width=9cm]{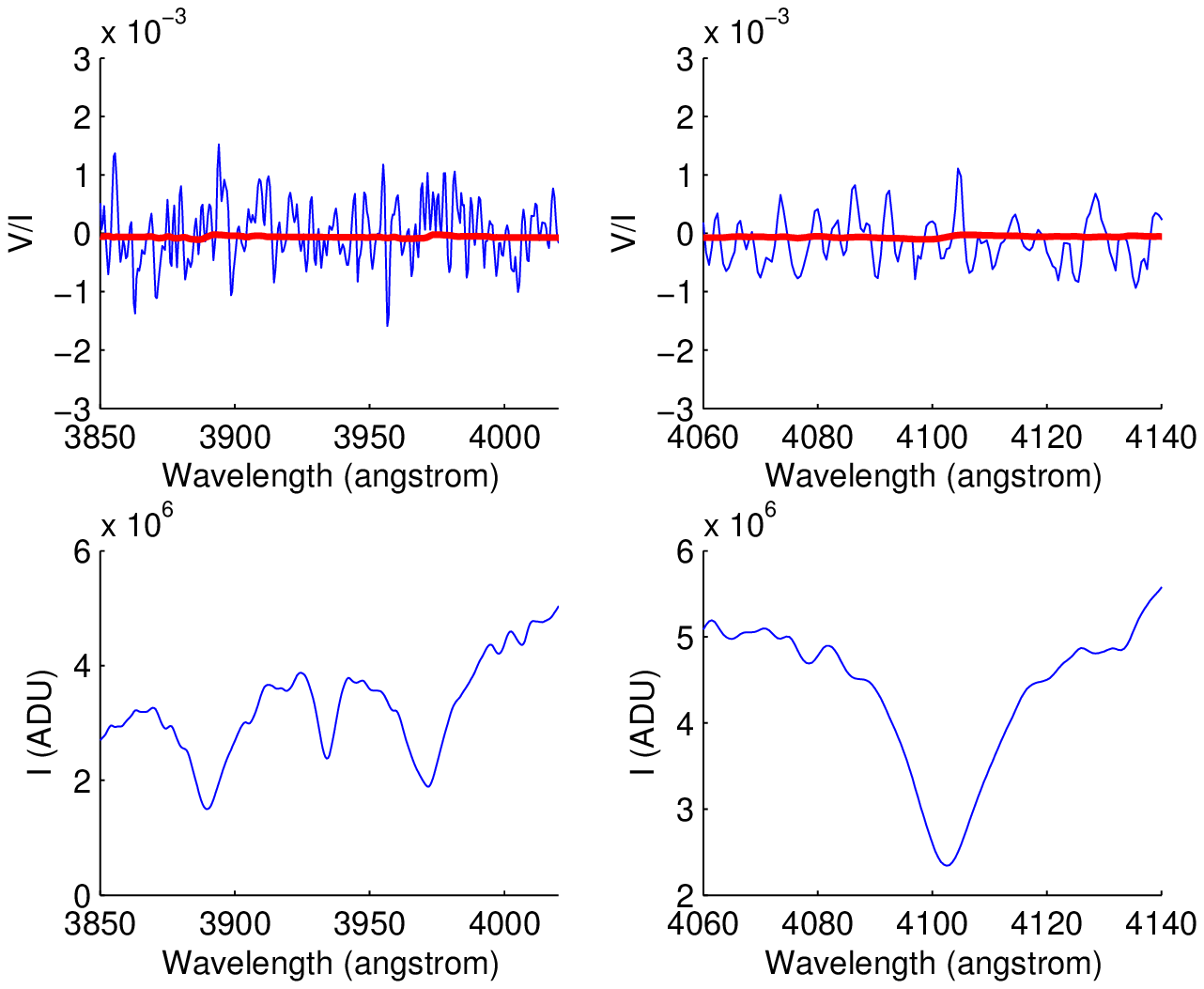}\includegraphics[width=9cm]{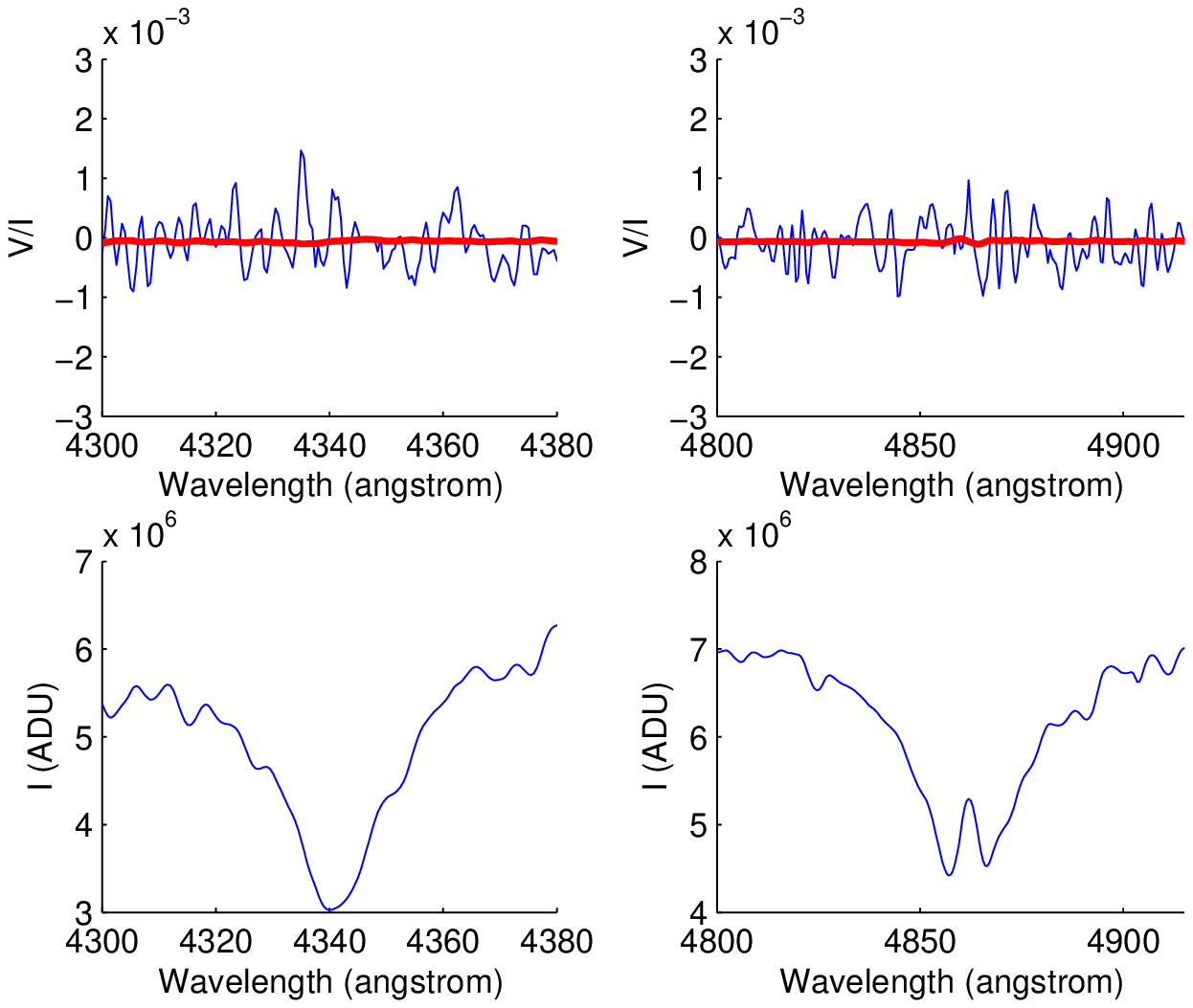}

\caption{Stokes $I$ and $V$ profiles in the spectrum of HD 36112. }

\end{figure*}

HD~36112 (PDS 183, MWC 758) is a well-studied HAeBe star, selected from the catalogue of Th\'e et al. (1994). Vieira et al. (2003) report H$\alpha$ and O~[{\sc i}] forbidden line emission. This star has been studied in detail by Beskrovnaya et al. (1999), who discuss the spectroscopic, photometric and polarimetric properties. They point out in particular the strong near-IR excess due to warm circumstellar dust, and peculiar, rapid variability of circumstellar line profiles, which they speculate may be connected with circumstellar jet-like inhomogeneities.

The Hipparcos parallax of $4.89\pm 1.16$~mas yields a distance of $205^{+63}_{-40}$~pc. In combination with the effective temperature $T_{\rm eff}=7700\pm 200$~K (Beskrovnaya et al. 1999) and the model evolutionary tracks of Palla \& Stahler (1993), this yields a mass of $1.8\pm 0.2$~$M_\odot$, and an age of 5-10 Myr (consistent with the results of Mannings \& Sargent (2000), who find 6 Myr). The projected rotational velocity of HD 36112 is $60\pm 6$~km/s (Beskrovnaya et al. 1999).

HD 36112 was observed during P74. Eight exposures were obtained, for a net S/N of 3725:1. The spectrum, illustrated in Fig. 7, is rich in metal lines and shows clear core emission at H$\beta$, and emission at He~{\sc i}~$\lambda 5876$. The Ca~{\sc ii}~K and Mg~{\sc ii}~$\lambda 4481$ lines are prominent. The magnetic diagnosis of this star (``nddn'') yields no detection in Balmer lines, but marginal detections according to one of the criteria of Bagnulo et al. (2006) in the metallic lines ($-166\pm 56$~G; $3.0\sigma$) and the full spectrum ($-149\pm 41$~G; $3.6\sigma$). Although this does not constitute strong evidence for a magnetic field, given the activity discussed by Beskrovnaya et al. (1999), this star should be reobserved.


\subsubsection{Other targets of note}

\vspace{5mm}
{\noindent \bf HD 72106A}
\vspace{5mm}






Wade et al. (2005) reported the detection of a weak longitudinal magnetic field in HD 72106A using both FORS1 and ESPaDOnS. The FORS1 detection, which corresponded to 5.2$\sigma$ (Balmer lines) and 5.8$\sigma$ (full spectrum) detections, was based on an earlier reduction and analysis of these data (Drouin 2005). The current, standardised reduction and analysis presented in this paper (according to the methodology reported by Bagnulo et al. 2006), which we believe to be more accurate and robust and to provide more realistic error bars, does not result in a field detection for this star (longitudinal fields of $166\pm 70$~G in Balmer lines and $-11\pm 91$~G in metal lines). {Nor is a magnetic field detected in HD 72106B (detection flag ``nnnn'', $\Bz=76\pm 67$~G from the full spectrum analysis). The difference in the results for HD 72106A can be explained as a natural result of the somewhat larger error bars derived here, coupled with the expected statistical fluctuations in measurements obtained from independent reductions of the data.} However, the magnetic field of HD~72106A was detected repeatedly by Wade et al. (2005) {and Folsom et al. (2006)} using the ESPaDOnS high-resolution spectropolarimeter, { with clear Stokes $V$ signatures and longitudinal field error bars of about 60~G}: HD 72106A is therefore a confirmed magnetic object. 

HD 72106B (PDS 031S - the fainter southern component of this $0.8\arcsec$\ double star; Harkopf et al. 1996; ESA 1997) was selected from the investigation of HAeBe group members and candidate stars by Vieira et al. (2003). According to those authors, this star exhibits weak H$\alpha$ emission and significant IR excess, and may be an evolved HAeBe object, possibly nearing the end of its PMS evolution. Both stars exhibit large proper motions and there is no detectable relative motion of the pair (Harkopf et al. 1996). {According to Folsom et al. (2006), the radial velocities are identical within the measurements uncertainties.} This suggests that they are physically associated, which implies that they are likely co-eval. We observed both stars individually, under conditions of excellent seeing, using a slit size of $0.5\arcsec$. 

Although no temperatures are available in the literature for the individual stars, based on the $B-V$ colours of Fabricius \& Marakov (2000) both components appear to the A-type stars. Our two FORS1 spectra confirm this, with well-developed Balmer lines and similar metallic-line spectra. The Balmer lines have similar depth in spectra of both stars, but they are considerably broader in the B component. The Mg~{\sc ii} $\lambda 4481$ lines are well-developed (stronger in the spectrum of the A component), as is the Ca~{\sc ii}~K line (substantially stronger in the B component). Overall, the A component's spectrum is similar to that of HD 94660 - the metallic spectrum is weaker, and the Balmer lines marginally broader. On the other hand, the B component's spectrum is suggestive of a cooler star. We have employed the FORS1 spectra of the two components, with their robust continua and Balmer line profiles, in combination with ATLAS9 Balmer line profiles (convolved to the same resolution as the FORS1 spectra assuming a gaussian instrumental profile), to constrain the temperatures of the two components. We derive from the comparison of observed and theoretical profiles that HD 72106A has $T_{\rm eff}=11000\pm 1000$~K, $3.5\leq \log g\leq 4.5$ and HD~72106B has $T_{\rm eff}=8000\pm 500$~K, $4.0\leq \log g\leq 4.5$. The best-fit results (for $T_{\rm eff}=11000$~K, $\log g=4.0$ and $T_{\rm eff}=8000$~K, $\log g=4.5$) are shown in Fig. 8.

Vieira et al. (2003) associated HD 72106 with the Vela and Gum Nebula SFRs. The system has a combined Hipparcos parallax of $3.47\pm 1.43$~mas, yielding a distance of $288^{+202}_{-84}$~pc, which is consistent with that of the SFR. Wade et al. (2005) estimated its position on the HR diagram. This position, interpreted using the model evolutionary tracks of Palla \& Stahler (1993), indicates that the components of HD~72106 have masses of $1.75\pm 0.25~M_\odot$ (B component) and $2.4\pm 0.4~M_\odot$ (A component), and are situated close to the ZAMS. {Although uncertain, these H-R diagram positions are consistent with a co-eval system.}

\begin{figure}

\centering

\includegraphics[width=8cm]{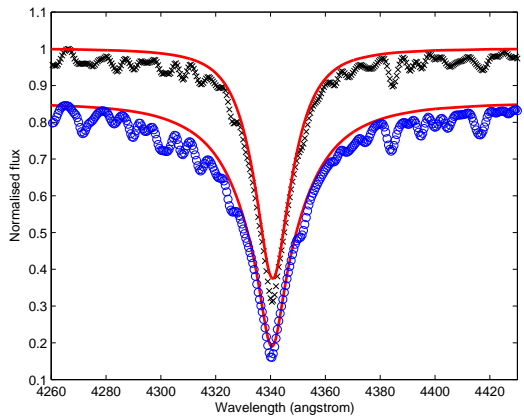}

\caption{Observed (symbols) and computed (lines) H$\gamma$ profiles of HD 72106A (brighter northern component - Top) and HD 72106B (fainter southern component - Bottom).}

\end{figure}

\vspace{5mm}
{\noindent \bf HD 144432, HD 31648 and HD 144668}
\vspace{5mm}

 HD 144432, HD~31648 and HD 144668 are 3 HAeBe stars that have been discussed by Hubrig et al. (2004, 2006a). These authors also used FORS1 to search for magnetic fields in Balmer lines, reporting marginal detections ($119\pm 38$~G ($3.1\sigma$) and $87\pm 22$~G ($4.0\sigma$) respectively) in HD~144432 and HD~31648, and no detection in HD~144668. {Of particular interest is the report by Hubrig et al. (2006b) of strong Stokes $V$ signatures in some spectral lines of these stars, which they interpret to be due to ``circumstellar'' magnetic fields. }

We have observed HD 144432, HD~31468 and HD 144668 as part of this programme. None of these stars yields any significant magnetic field detection in their metallic or Balmer lines, with uncertainties comparable to those of Hubrig et al. (2004). {We have also carefully examined our reduced spectra, and we find no evidence of strong Stokes $V$ signatures in our observations of these stars. }













\begin{figure*}

\centering

\includegraphics[width=9cm]{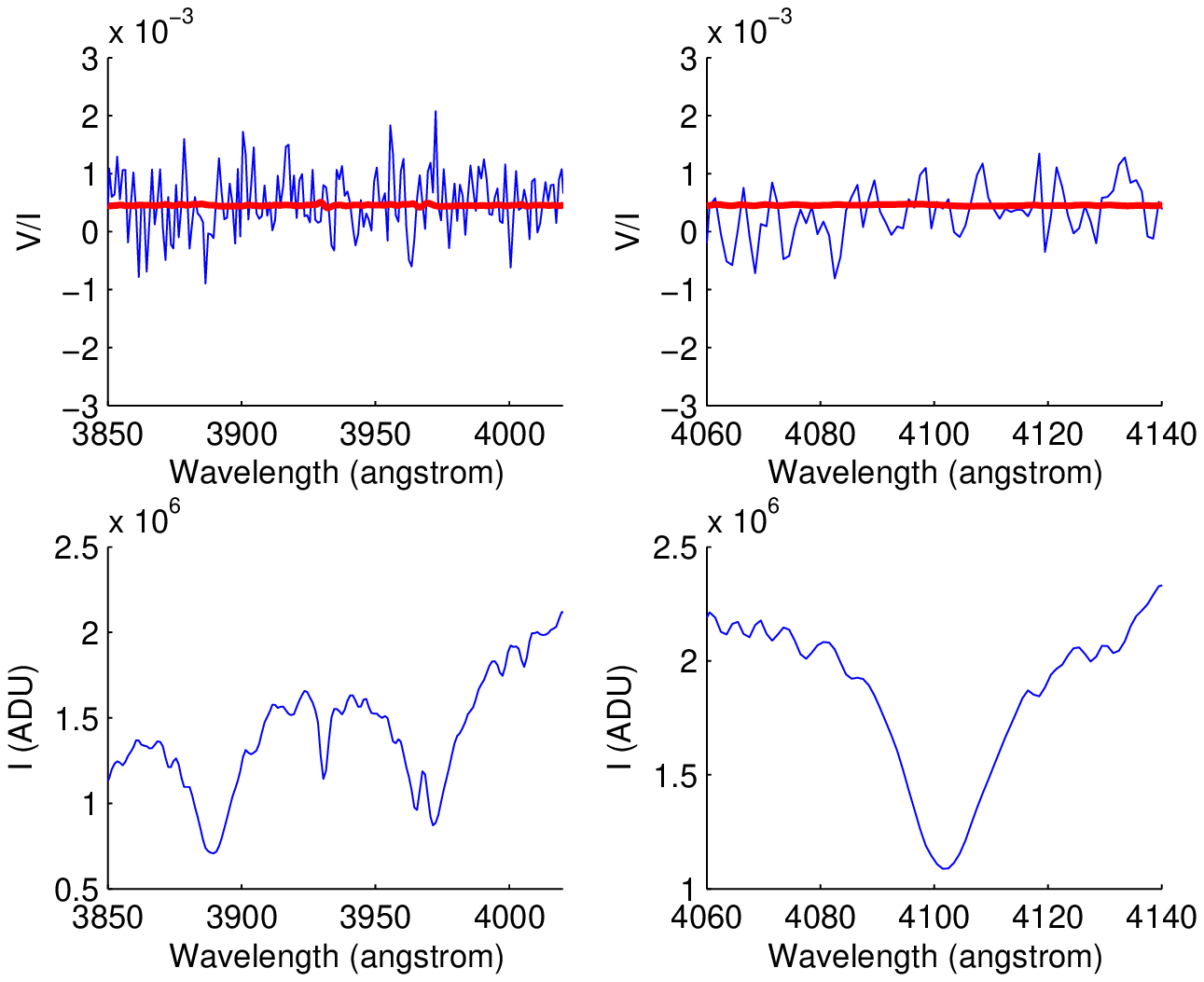}\includegraphics[width=9cm]{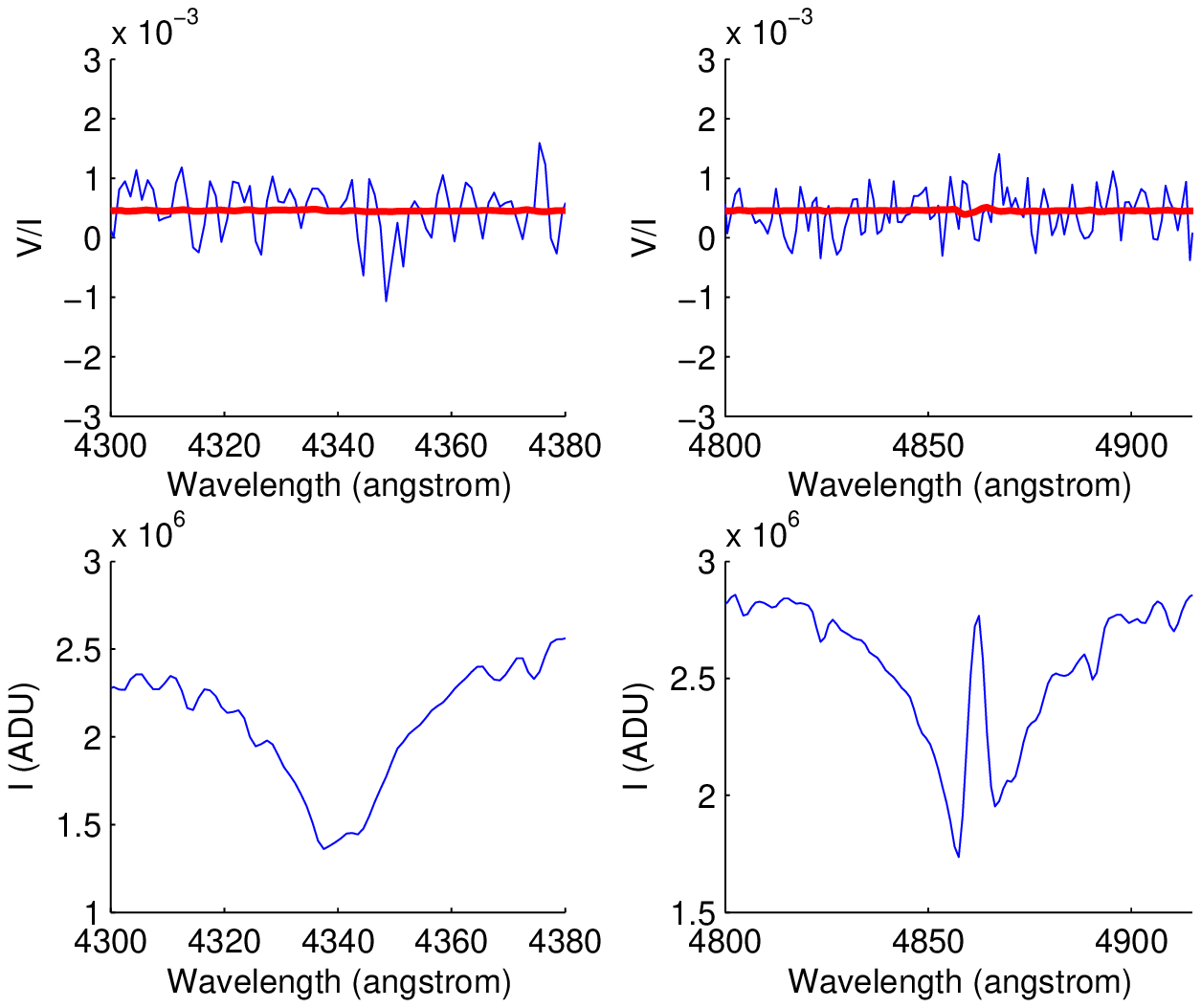}

\caption{Stokes $I$ and $V$ profiles in the spectrum of HD 104237.}

\end{figure*}

\vspace{5mm}
{\noindent \bf HD~104237}
\vspace{5mm}

HD 104237 is a cool Herbig Ae star with a sharp metallic-line spectrum in which Donati et al. (1997) claimed the marginal detection of a magnetic field using high-resolution spectropolarimetry and Least-Squares Deconvolution. This result was subsequently confirmed by Donati (2000), {who detected Stokes $V$ signatures and derived a longitudinal field of about 50~G (Donati, priv. comm.)} using new data and similar techniques. Acke \& Waelkens (2004) found approximately solar photospheric abundances based on an equivalent width abundances analysis.

{ HD~104237 was observed during both P72 and P74. Three observations (corresponding to 12, 10 and 8 exposures) were obtained, for net S/Ns of 2375:1, 2800:1 and 2400:1. The spectrum, illustrated in Fig. 10, exhibits significant emission at H$\beta$, which shows a strong and variable emission core, as well as emission in the multiplet 42 Fe~{\sc ii} lines.} The magnetic diagnoses of this star yield no detection in either Balmer or metal lines according to any of the criteria of Bagnulo et al. (2006), with { best error bars of 64~G (Balmer lines) and 35~G (metal lines).} The uncertainties obtained here are not sufficiently small to detect a field of longitudinal intensity similar to that observed by Donati (2000). 


\vspace{5mm}
{\noindent \bf V380 Ori}
\vspace{5mm}

V380~Ori is a young B9 HAeBe star exhibiting numerous emission lines in its optical spectrum, associated with essentially all metallic, H and He transitions. High resolution spectra (Wade et al. 2005) reveal that many lines have a weak, sharp absorption component which appears to be at least partly photospheric in nature {(e.g. as suggested by the successful Least-Squares Deconvolution analysis by Alecian et al. 2006)}. Rossi et al. (1999) have examined the photometric and spectroscopic properties of this star, concluding that it hosts a Keplerian disc and a cool wind. Leinert et al. (1997) discovered that V380 Ori is an interferometric binary, with a companion that is 3-4 times fainter in K band, has $v\sin i\sim 30$~km/s and which is sufficiently cool to display the Li~{\sc i}~$\lambda$6708 line in its spectrum (Corporon \& Lagrange 1999).

V380~Ori was observed during P74. Four exposures were obtained, for a net S/N of 1450:1. The spectrum, illustrated in Fig. 11, exhibits signicant emission in nearly all lines, including higher Balmer series members. The magnetic diagnosis of this star yields no detection in either Balmer or metal lines according to any of the criteria of Bagnulo et al. (2006), with 1$\sigma$ uncertainties of 66 G in Balmer lines and 24 G in metal lines. Although undetected in this study, this star has been detected twice with ESPaDOnS by Wade et al. (2005), with longitudinal fields derived from the detected Stokes $V$ signatures of $29\pm 35$~G and $460\pm 70$~G. {Recently, Alecian et al. (2006) performed a preliminary analysis of the ESPaDOnS Stokes $V$ timeseries, deriving a surface dipole magnetic field of 1.4~kG}. Examination of Fig. 1 of Wade et al. (2005) shows that the circular polarisation signatures they detect are associated with the weak absorption lines superimposed on the strong emission lines. It seems likely that the polarisation would not be detectable at the low resolution of FORS1, due to the strong influence of unpolarised emission in each line.

\begin{figure*}

\centering

\includegraphics[width=9cm]{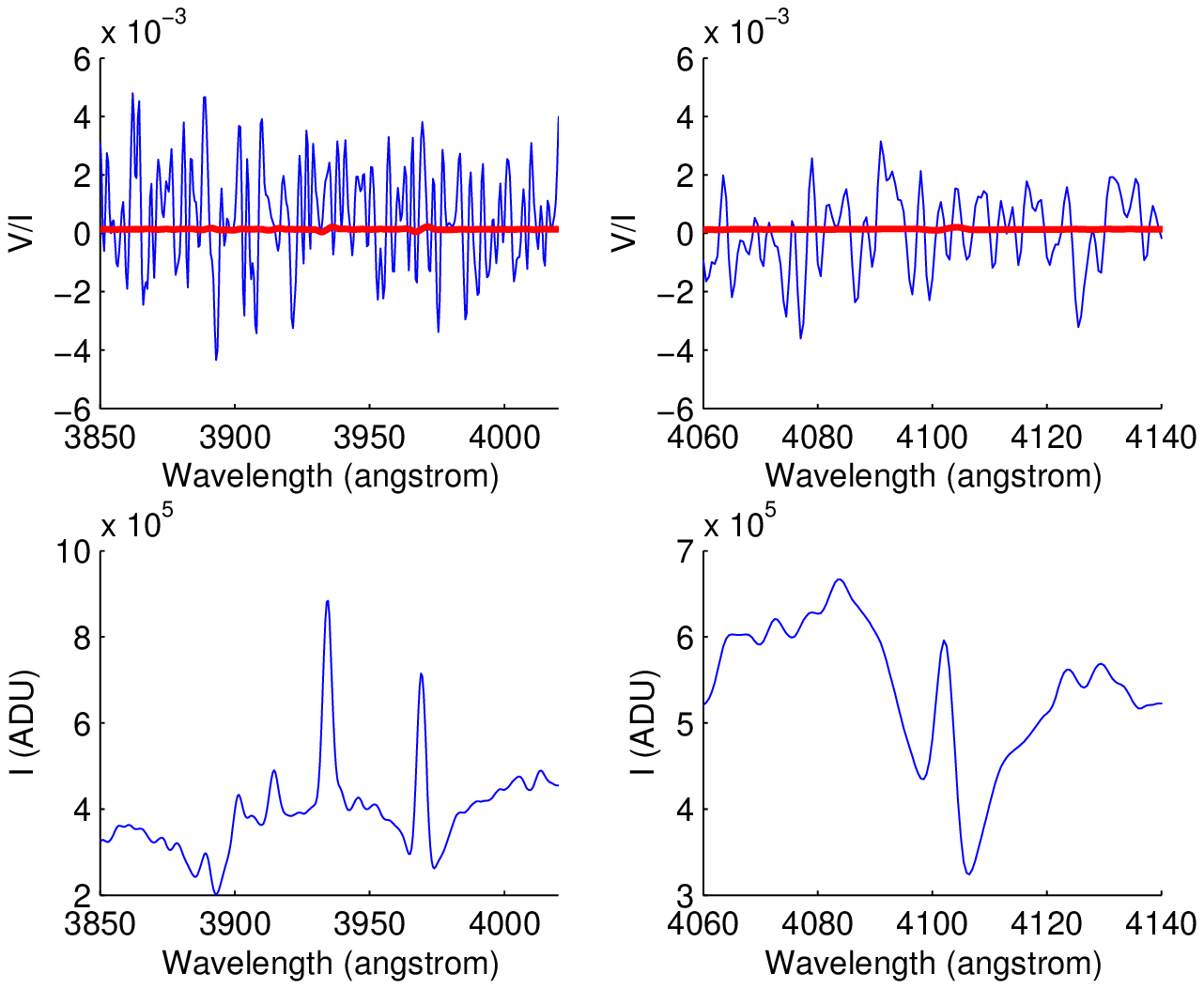}\includegraphics[width=9cm]{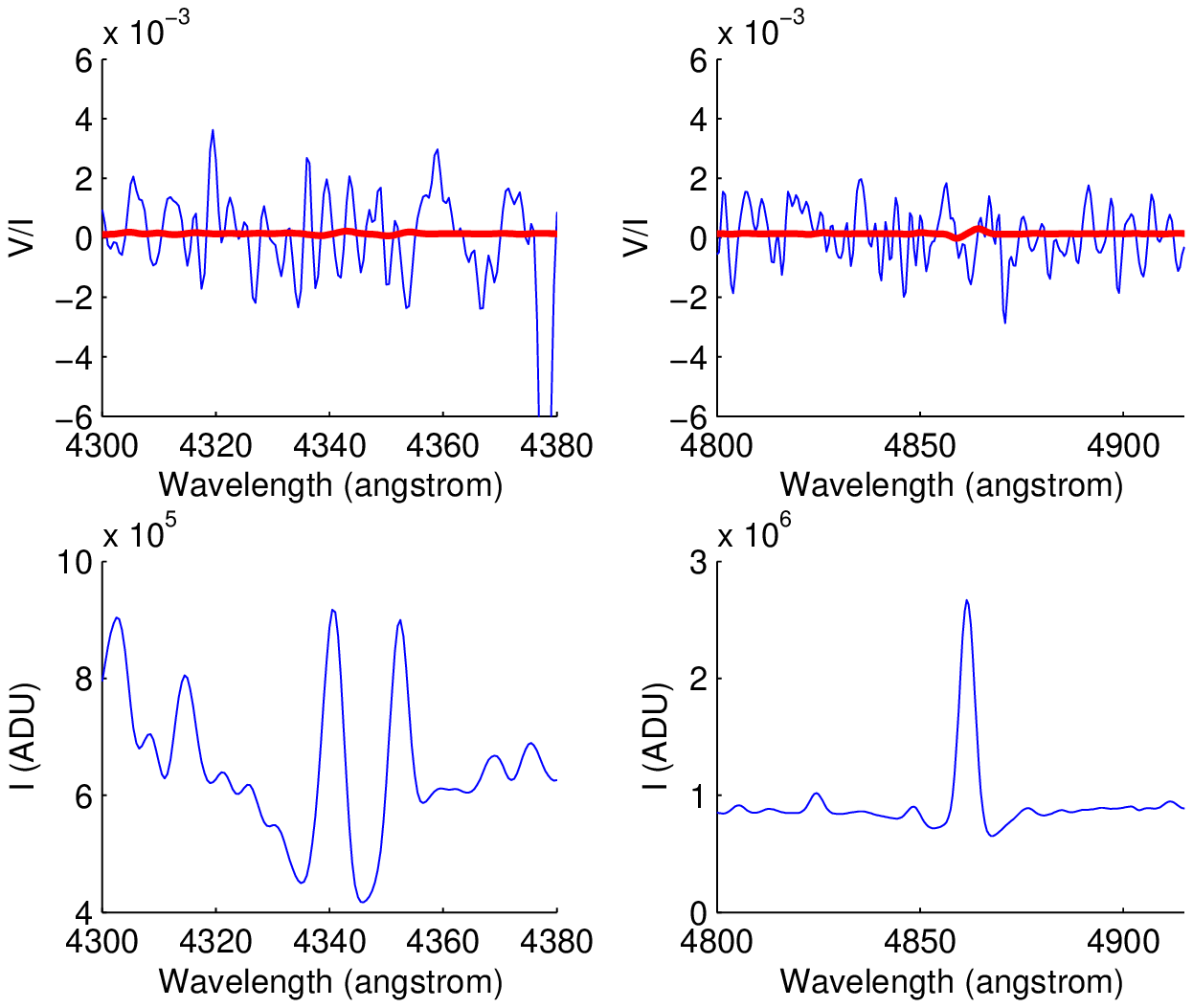}

\caption{Stokes $I$ and $V$ profiles in the spectrum of V380 Ori. }

\end{figure*}

\vspace{5mm}
{\noindent \bf HD~190073}
\vspace{5mm}

HD~190073 is an A2 HAeBe star, showing a spectrum with a large number of emission lines. Like V380 Ori, many emission lines have superimposed narrow absorption
cores, with a width similar to the photospheric lines with no emission. As reported by Catala et al. (2006), the emission components have a uniform width equal to 65~km/s. P Cyg profiles are exhibited by H$\alpha$ and other Balmer lines, with strong monthly variability. Acke \& Waelkens (2004) report no significant chemical peculiarity (Ca may be marginally enhanced). Pogodin et al. (2005) have reported complex structure in the Ca~{\sc ii}~H and K line profiles, and suggest that a magnetic field could account for the particular circumstellar structure of this star.

HD~190073 was observed during P74. Two observations were obtained { (one with grism 600B and one with grism 1200g)}, each comprised of 8 exposures, yielding S/Ns of 3700:1 and 3200:1. The magnetic diagnoses of this star yield no detection in either Balmer or metal lines according to any of the criteria of Bagnulo et al. (2006). Although undetected in this study, circular polarisation has been detected repeatedly in the photospheric metal lines of this star with ESPaDOnS (Catala et al. 2006), who consistently observe Stokes $V$ profiles which do not vary detectably on a one-year timeframe, indicating either an azimuthally symmetric photospheric magnetic field, a zero inclination angle between the rotation axis and the line of sight, or a very long rotational period. As with V380~Ori, we attribute the lack of detection with FORS1 to the weakness of the field (below 100~G (Catala et al. 2006) {and therefore too weak to produce detectable Stokes $V$ signatures in our spectra}) and the unpolarised contributions of the many emission lines. 

{As with HD 144432, HD 31648 and HD 144668, Hubrig et al. (2006b) observe strong Stokes $V$ signatures in some spectral lines of this star, which they claim to be due to ``circumstellar'' magnetic fields. We do not confirm the detection of such features in our spectra. As a further test, to check if we could confirm this strong polarisation in the same data employed by Hubrig et al., we have extracted from the ESO Science Archive the 3 spectra of HD 190073 discussed by Hubrig et al. (2006b). We reduced and analysed the data using the procedures of Bagnulo et al. (2006), then examined the spectra in detail. We find no evidence for the frequent presence of strong, systematic polarisation as reported by those authors. In one spectrum, we do observe a marginal spike in Stokes $V$ associated with the absorption feature near the position of Ca~{\sc ii}~K. However, the width of this spike (1.2~\AA) is below the limit of the resolving power (the dispersion is just 0.6~\AA}). Hence it has little value for physical interpretation.} 


\section{Discussion}

\subsection{Survey summary}

 We have obtained measurements of the longitudinal magnetic fields of 50 Herbig Ae/Be stars, obtaining typical 1$\sigma$ error bars of 66~G (Balmer lines) and 105~G (metal lines). The longitudinal field is derived using 4 subsets of spectral lines and 2 complementary methods for obtaining the final longitudinal field intensity, for a total of 272 field measurements { (not all of which are independent)}. Twelve of these measurements correspond to formal detections of a nonzero longitudinal field ($|\bz|\geq 3\sigma_B$), ultimately yielding {4} stars (HD 101412, BF Ori, {CPD-53 295} and HD 36112) for which a magnetic field is possibly detected.

As discussed in Sect. 4, none of these stars qualifies as a definite field detection (diagnosis flag ``DDDD'') according to the criteria of Bagnulo et al. (2006). Because none of the stars exhibit really strong fields, we expect the metallic-line and Balmer line magnetic field diagnoses to be consistent (Bagnulo et al. 2006). Although emission in Balmer lines may cause a breakdown of this consistency for HAeBe stars, none of the 4 stars show strong emission in higher members of the Balmer series (H$\gamma$ onward). We have concluded that HD 101412 is probably magnetic, obtaining a longitudinal field (obtained from the full spectrum) of $446\pm 106$~G. We also suspect BF Ori to be magnetic, but with less confidence due to the weaker inferred field ($-144\pm 21$~G). { CPD-53 295, showing magnetic field only in metallic lines and in only one of two observations obtained, and HD 36112, also showing only a marginal detection in metallic lines, are less probably magnetic,} and we consider them non-magnetic stars for the remainder of this analysis. In any case, all of these stars should be re-observed to verify the presence of magnetic fields.

We therefore tentatively conclude that 2 new magnetic HAeBe stars are detected in this survey. Taking into account that the longitudinal field of a dipole magnetic field is never larger than about $1/3$ of the dipole polar intensity, we can estimate {\em lower limits} on the surface dipole intensities of these stars: $1.3\pm 0.3$~kG (HD 101412) and $0.43\pm 0.06$~kG (BF Ori). {More highly-structured fields (e.g. higher-order multipoles) would require even larger surface intensities per unit longitudinal field. 

For the remainder of the stars in this study, our null results allow us to place constraints on the general magnetic properties of HAeBe stars. This is discussed in detail later in this section.}

\subsection{Other published results}

{In addition to the previously published results already discussed in Sect. 4 (HD 72106 and V380 Ori (Wade et al. (2005), HD 144432, HD 31648 and HD 144668 (Hubrig et al. 2004, 2006a), HD 104237 (Donati et al. 2000), and HD 190073 (Catala et al. 2006)), some additional recent magnetic data for other HAeBe stars is also available in the literature.

Alecian et al. (2006) demonstrated the repeated detection of Stokes $V$ signatures in 21 observations of the Herbig Be star HD~200775, which they successfully reproduced assuming an oblique, rotating magnetic dipole with a polar strength of 400~G. Although HD~200775 has not been observed in our study, the repeated detection of variable Stokes $V$ signatures that were coherently interpreted using a simple magnetic model strongly supports the reality of the detected field.

Hubrig et al. (2004, 2006a) reported two FORS1 detections of a longitudinal magnetic field in HD~139614. However, Wade et al. (2005) showed that high-resolution Stokes $V$ LSD profiles of this cool, sharp-lined Herbig Ae star, obtained on multiple nights, reveal no evidence of a magnetic field. Although Hubrig et al. (2006b) dismiss this result, supposing that LSD is inappropriate for magnetic field diagnosis of HAeBe stars, the results reported by Alecian et al. (2006), Folsom et al. (2006) and Catala et al. (2006) demonstrate that LSD can be used effectively for field diagnosis in HAeBe stars, even those stars with substantially more complex spectra than HD~139614. Hence it appears that credible evidence is lacking for the presence of a detectable magnetic field in HD 139614.

}  

\subsection{General magnetic properties of Herbig Ae/Be stars}

{For the large majority of stars in this study in which we do not detect magnetic fields, our null results allow us to generally rule out the presence of ordered magnetic fields frequently producing longitudinal fields stronger than about $3\sigma_B$. For a rotating magnetic dipole titled at an angle $\beta$ with respect to the stellar rotation axis, the longitudinal field varies sinusoidally according to the stellar rotation, with a mean intensity and amplitude that are functions of the inclination of the stellar rotational axis to the observer's line-of-sight $i$, the inclination of the dipole axis to the rotation axis $\beta$, and the dipole polar strength $B_{\rm d}$. When $i=0$ or $\beta=0$, the longitudinal field does not vary (i.e. it is constant during the stellar rotation). We have attempted to model the observed detection significance distribution (shown in Fig. 4) for the Balmer line measurements using a Monte Carlo scheme, computing synthetic values of $z=\bz/\sigma_B$ assuming a dipolar magnetic field. In constructing the model distributions, we have assumed a random distribution of $i$, we have sampled each longitudinal field variation at a random phase (i.e. we assumed a single observation of each star), and we have assigned observational errors and gaussian noise consistent with the distribution shown in Fig. 3. Comparison with a range of model assumptions for unifom populations yields the following conclusions:

\begin{enumerate}

\item Non-magnetic population: Excluding the measurements of BF Ori and HD 101412, the observed detection significance distribution is consistent within statistical uncertainty with a distribution of non-magnetic stars (i.e. a uniform distribution of $B_{\rm d}=0$).
\item Uniform population of aligned rotators: Excluding the measurements of BF Ori and HD 101412, the observed detection significance distribution is inconsistent with a uniform population of magnetic stars with aligned magnetic fields ($\beta=0$), if $B_{\rm d}\gtrsim 300$~G.
\item Uniform population of oblique rotators: Excluding the measurements of BF Ori and HD 101412, the observed detection significance distribution is inconsistent with a uniform population of magnetic stars with perpendicular magnetic fields ($\beta=90\degr$), if $B_{\rm d}\gtrsim 500$~G. 

\end{enumerate}

For multicomponent populations (e.g. a combined population of magnetic and non-magnetic stars), the upper limits are generally increased. For example, a equal mix of non-magnetic and magnetic aligned rotators yields an upper limit for $B_{\rm d}$ of about 500~G, while a similar mix of perpendicular rotators yields an upper limit of about 750~G.}

\subsection{Magnetic properties of HAeBe stars vs. main sequence A and B stars: testing the fossil-field hypothesis}

 {The primordial fossil-field hypothesis, which proposes that the magnetic fields of Ap/Bp stars are the slowly-decaying remnants of interstellar magnetic field swept up during star formation, requires that the magnetic properties of HAeBe stars be qualitatively the same as those of the main sequence A and B stars. Here we compare the known magnetic properties of the main sequence A and B type stars with our observations of HAeBe stars.

 We begin by comparing the complete observed detection significance distribution with another Monte Carlo simulation: a model combined population composed of 1~kG perpendicular rotators (10\% of the population) and non-magnetic stars (90\% of the population). This sample roughly approximates the main sequence intermediate-mass population, including the Ap/Bp stars. The observed and computed distributions are statistically identical, demonstrating that our observations of HAeBe stars are consistent with a magnetic field distribution similar to that of the main sequence stars. This is consistent with the fossil-field hypothesis. However, as the observed distribution is also consistent with a uniform population of non-magnetic stars, this comparison only demonstrates that our sample of HAeBe stars does not contain a significantly larger number of stars with strong fields than the main sequence stars. }

 In addition to the 2 stars in which we tentatively claim new field detections, four other stars in our sample (HD 104237, HD 72106A, V380 Ori and HD 190073) have been detected and confirmed previously by other authors (Donati et al. 1997; Wade et al. 2005; Catala et al. 2006), but are not detected in this study (for reasons discussed in Sect. 4). Thus, the total number of confirmed magnetic stars in this sample is 4, and the number of suspected magnetic stars is 2. This corresponds to a bulk incidence of (detected) magnetic stars of between 8\% and 12\% (taking into account the possibility that HD 101412 and BF Ori may not be magnetic). This range is fully consistent with the canonical incidence of 5-10\% (e.g. Wolff 1968) of magnetic Ap and Bp stars on the main sequence, and therefore with the fossil-field hypothesis. In fact, because the observed sample is strongly biased to stars with masses from $2-3~M_\odot$, which correspond to approximate main sequence $B-V$ colour indices of +0.14 to -0.1, we would expect that the magnetic field incidence of a sample of main sequence stars with a similar distribution of masses to be somewhat higher than the canonical value (based on the incidence distributions of Wolff 1968).

\begin{figure*}

\centering

\includegraphics[width=9cm]{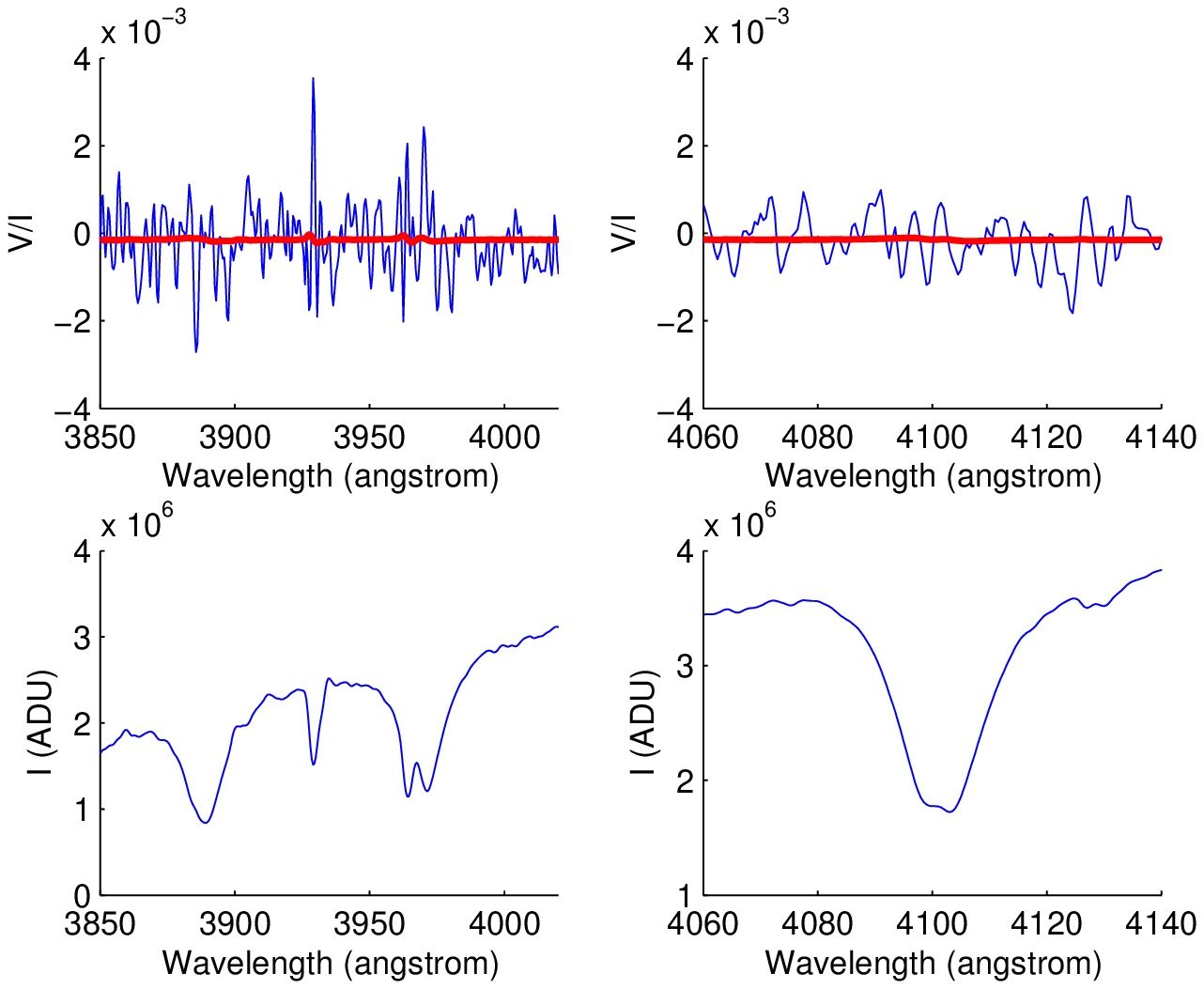}\includegraphics[width=9cm]{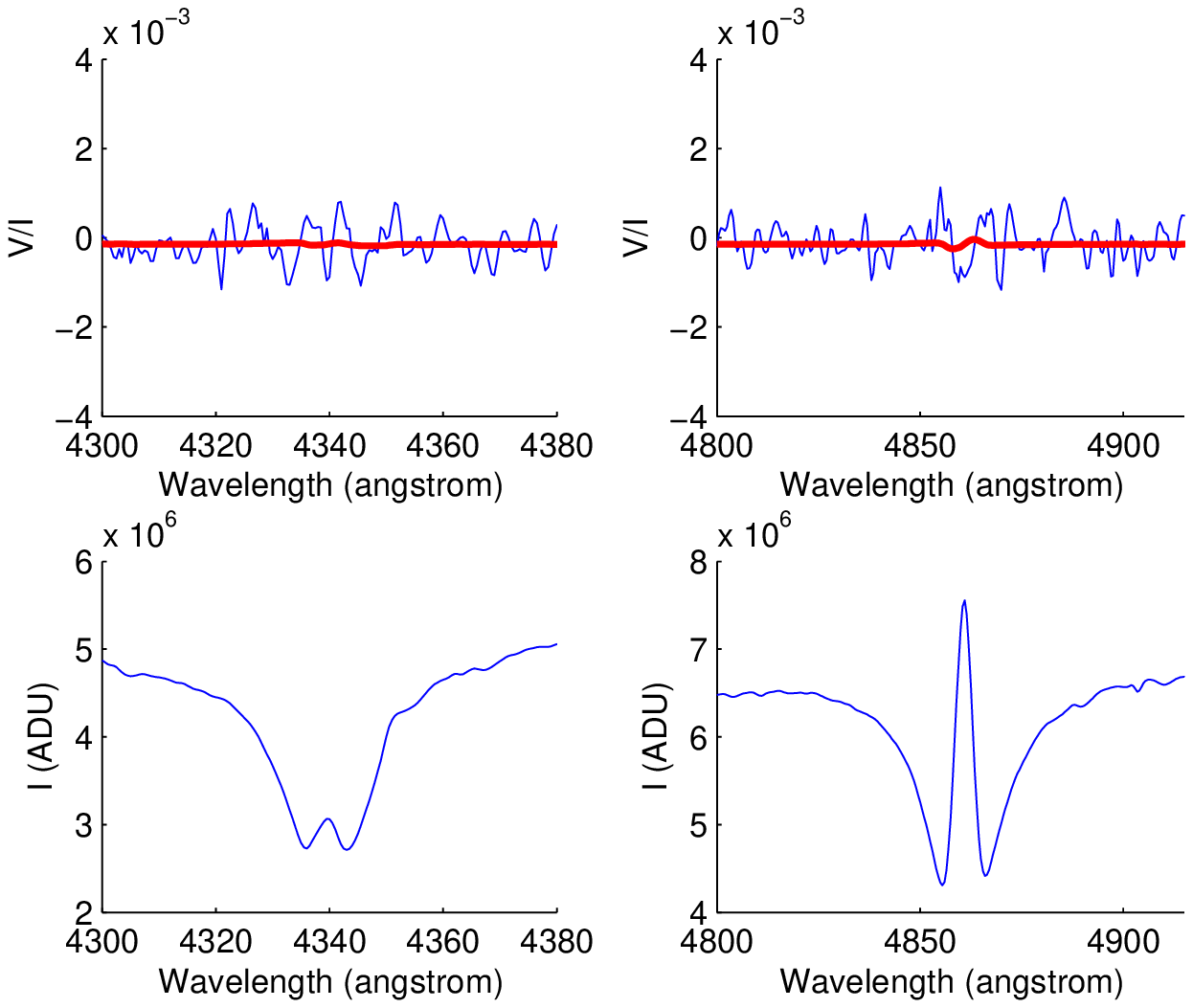}

\caption{Stokes $I$ and $V$ profiles in the spectrum of HD 190073.}

\end{figure*}

Under the fossil-field hypothesis, the magnetic fields of the detected Herbig Ae/Be stars are to become those of Ap/Bp stars on the main sequence. We should therefore expect similar magnetic field geometries in both classes of stars. Unfortunately, because we have only a single longitudinal magnetic field observation of most of the studied stars, it is difficult to constrain the geometry of their magnetic fields. However, the results reported by Wade et al. (2005), {Folsom et al. (2006) and Alecian et al. (2006) for HD~72106A, V380 Ori and HD~200775, and by Catala et al. (2006) for HD~190073, indicate that the magnetic fields of these objects are clearly organised on large scales, and that the fields of V380 Ori, HD~72106A and HD~200775 have important dipolar components.} The strong longitudinal fields measured here for HD~101412 and BF~Ori also suggest that these fields are organised on large scales - longitudinal fields corresponding to the more complex magnetic topologies associated with dynamo-generated fields {in cool stars} seldom exceed a few tens of G for stars of moderate activity, and peak at about 100~G for the most active late-type stars (Petit et al. 2005; Petit, private communication). It therefore appears likely that for most, and possibly all, of the detected stars have magnetic fields that are structured on global scales. This is consistent with the fossil-field hypothesis.

 Bohlender \& Landstreet (1990) examined the longitudinal magnetic field characteristics of the sample of 12 Ap stars brighter than $V=5.0$ north of $-60\degr$. {Although small, this magnitude-limited sample probably provides a realistic approximation of the real distribution of most magnetic field strengths of Ap stars\footnote{In particular, Ap stars with very strong magnetic fields (e.g. exhibiting longitudinal fields more than 1~kG) are extremely rare, and form only the extreme tail of a distribution in which the large majority of stars have much weaker fields.}.} They found that the {typical} rms longitudinal magnetic field of the sample is 330~G, with {an intrinsic spread} of about 170~G. The mean rms longitudinal field measured for the HAeBe stars detected in this and other studies is roughly 200~G (Donati et al. 1997 and Donati, priv. comm.; Wade et al. 2005; Catala et al. 2006; Alecian, priv. comm.). The uncertainty on the latter value is difficult to quantify, given the sparse observations and the diversity of data sources, but it is probably of order 100~G. 

If the fossil-field model is indeed correct, we should in fact expect the fields of HAeBe stars to be somewhat weaker than those of the Ap stars, due to the larger radii of the PMS stars. This ``field dilution'' can be calculated approximately for each star using the squared ratio of the current stellar radius to the radius of that star upon reaching the ZAMS, $(R_{\rm PMS}/R_{\rm ZAMS})^2$. This was performed using the interpolated PMS evolutionary model calculations of Palla \& Stahler (1993). Each star in Fig. 1 was associated with an evolutionary track and ``evolved forward'' to the ZAMS, at which point its radius was calculated from its predicted ZAMS luminosity and effective temperature. The average ``field dilution'' factor is about 0.5, indicating that we should expect the mean rms longitudinal field of HAeBe stars in this sample to be about 50\% weaker than that of ZAMS Ap/Bp stars, if the latter descend from the former. {In fact, the sample of Ap stars studied by Bohelender \& Landstreet (1994) is probably somewhat evolved; hence the calculated dilution probably represents an upper limit on what we could expect to observe. {Utimately, given the large intrinsic spread of the Ap stars and the uncertainties associated with the HAeBe stars, we are unable to critically constrain field dilution, and we note that rms field strengths of both groups of stars are approximately equal.}}

Because the magnetic fields of HAeBe stars are {expected, under such a model,} to be systematically weaker than those of Ap/Bp stars, the likelihood of obtaining observational ``false negatives'' is higher, in particular as we have no obvious way of pre-selecting (such as using chemical peculiarity, for example) {magnetic HAeBe stars from the general population of HAeBe stars.} It may therefore be that magnetic HAeBe stars exist in our sample that we have not detected. The current study is generally insensitive to longitudinal magnetic fields below about 150~G, and this sensitivity limit is worse for stars with significant {contamination of metallic or Balmer lines by emission} from circumstellar material. In addition, this study is unable to directly rule out the general presence of magnetic fields with complex structure (e.g. dynamo-generated fields) in the Herbig Ae/Be stars: as discussed previously, the maximum longitudinal field intensity of dynamo-type magnetic fields is below our general detection threshold. {Ultimately, to the extent that these data are able to test the fossil-field hypothesis, they are in full agreement with its principles.}

\subsection{Comments on magnetospheric accretion in Herbig Ae/Be stars}

{As was discussed in Sect. 1, a motivation of this study is to explore the general role of magnetic fields in the late stages of formation of intermediate-mass stars. Some investigators believe that HAeBe stars are the higher mass analogues of classical T Tauri stars (CTTS), and that all of the phenomena associated with CTTS are also operating in HAeBe stars, including magnetospheric accretion (see for example Muzerolle et al. 2004).  

Magnetospheric accretion requires the presence of strong, large-scale magnetic fields at the stellar surface, and in the case of CTTS the predicted field strengths range up to several kG for specific stars. Johns-Krull et al. (1999) review three magnetospheric accretion models, providing equations yielding the surface magnetic field intensity required by each. These models assume aligned magnetic dipoles.
 
We can use these same equations to roughly estimate the field strengths required for magnetospheric accretion in HAeBe stars. According to Table 1, the mean measured $v\sin i$ of our sample is 115~\kms. For a HAeBe star with a radius of 2~$R_\odot$ (and assuming a rotational axis inclination $i=90\degr$ for simplicity), this implies a rotational period of about 1 day. Using the 3 models described by Johns-Krull et al., for a HAeBe star with mass $M=2~M_\odot$, radius $R=2~R_\odot$, rotational period $P=1$~day, mass accretion rate $M=10^{-8} M_\odot$/yr (see e.g. Garcia Lopez et al. 2006, Muzzerole et al. 2004), and canonical values of model parameters $\alpha_x, \beta$ and $\gamma$, we compute that a dipole magnetic field with a {\em surface polar} intensity of about 500~G is required for magnetospheric accretion according to the models of K\"onigl (1991) and Shu et al. (1994), and about 100~G for the model of Cameron \& Campbell (1993). For all models, the required field strength increases with mass, period and accretion rate, and decreases with radius. 

{We have already reported that our observations are not consistent with a uniform distribution of aligned magnetic rotators with $B_{\rm d}\gtrsim 300$~G. Our data are therefore not consistent with the general presence of magnetic fields with the strengths required by the models of K\"onigl (1991) and Shu et al. (1994). We therefore conclude that magnetospheric accretion according to the recipes of K\"onigl (1991) and Shu et al. (1994) is not generally occurring in HAeBe stars. On the other hand, we are unable to rule out the general presence of the weaker fields required by the model of Cameron \& Campbell (1993).

Vink et al. (2002) report systematic differences in the linear polarisation signatures across the H$\alpha$ line in spectra of Herbig Ae versus Herbig Be stars. They tentatively propose that this may indicate a transition from magnetically-mediated accretion in the Ae stars to disc accretion in the Be stars. Our sample contains 14 late A-type stars (with spectral type A3 or later), and a similar number of B-type stars (13 with spectral type B9.5 or earlier). Excluding HD~101412 and BF~Ori (our tentative detections), we find no statistically significant difference between the detection significance distributions of these two samples. Therefore our observations provide no support for the suggestion of Vink et al. (2002).

For those HAeBe stars in which magnetic fields have been detected, magnetospheric accretion according to any of these models may still be a viable hypothesis. The plausibility of such a proposal can only be tested by detailed investigation of the magnetic and physical properties of individual stars. Such investigation is outside the scope of this paper, but it will form the focus of future work.}


 
}

\subsection{Concluding comments}

In conclusion, we comment that the general effectiveness of FORS1 for studies of the magnetic properties of Herbig Ae/Be stars appears to be rather limited. In support of this, we point out the lack of detection of 4 known magnetic stars in our sample (HD~104237, HD~72106A, V380 Ori and HD~190073), and to the questionable meaning of Eq. (1) when applied to Balmer lines in emission. Although FORS1 does in principle provide the ability to detect magnetic fields in very rapidly-rotating HAeBe stars for which high-resolution spectropolarimetry is less sensitive, its typical detection threshold in this study (about 150~G), apparent ambiguity of detection of fields with longitudinal intensities close to this threshold, and its insensitivity to magnetic fields in stars with {emission-contaminated} line profiles are { substantial} handicaps, making it poorly competitive { for magnetic investigations of HAeBe stars} as compared to high-resolution spectropolarimeters such as ESPaDOnS at the Canada-France-Hawaii Telescope and the new NARVAL instrument, installed at the T\'elescope Bernard Lyot at Pic du Midi Obsevratory. {Nevertheless, the results of the present paper are of considerable interest. In particular, we have (1) made possible a first assessment of the field strengths present in HAeBe stars, {demonstrating the general absence of strong (0.5~kG), ordered magnetic fields}; (2) identified from a rather large sample a few apparently magnetic stars of great interest for further investigation; and (3) made possible a first estimate of the fraction of HAeBe stars which may evolve to become magnetic Ap/Bp stars. {All of these results are consistent with the primordial fossil field hypothesis, which proposes that magnetic fields in main sequence intermediate-mass stars are the slowly-decaying remnants of interstellar magnetic field swept up during star formation. Finally, we have performed a first evaluation of the plausibility of magnetospheric accretion for Herbig Ae/Be stars, ruling out the general applicability of the models of K\"onigl (1991) and Shu et al. (1994).}}

\section*{Acknowledgments}
GAW and JDL acknowledge Discovery Grant support from the Natural Science and Engineering Research Council of Canada (NSERC).

\begin{small}
\begin{longtable}{lcccccccccc}
\caption{Journal of observations. The table includes Julian date ($J2000$), the width of the slit (in arcsec), the exposre time and the number of exposures, and the magnetic field measurements. $B^{(1)}_z$ corresponds to the field measured in Balmer lines, $B^{(2)}_z$ corresponds to the field measured in metal lines, $B^{(3)}_z$ coorresponds to the field measured from the full spectrum, and $B^{(4)}_z$ corresponds to the field measured in all Balmer lines, excluding H$\beta$. S/N is the signal-to-noise ratio. Observations obtained using grism 1200g+96 are marked with an asterisk ($^*$). }\\
\hline\hline
Name & Julian Date & Slit width & Exp. Time & S/N & $B^{(1)}_z$ & $B^{(2)}_z$ & $B^{(3)}_z$ & $B^{(4)}_z$ & Flag \\
& & (arcsec) & (s) & & (Gauss) & (Gauss) & (Gauss) & (Gauss) & \\\hline
\endfirsthead
\caption{continued.}\\
\hline\hline
Name & Julian Date & Slit width & Exp. Time & S/N & $B^{(1)}_z$ & $B^{(2)}_z$ & $B^{(3)}_z$ & $B^{(4)}_z$ & Flag \\
& & (arcsec) & (s) & (Gauss) & (Gauss) & (Gauss) & (Gauss) & \\\hline
\endhead
\hline
\endfoot
HD 76534         &2453062.7192   & 1.0 & 940 (30)  & 4400 &   -39 $\pm$   75 &    58 $\pm$   64 &    55 $\pm$   47 &    12 $\pm$   81 &nnnn\\
HD 94660         &2453062.7611   & 0.5 & 24 (8)    & 1900 & -2534 $\pm$   63 & -1617 $\pm$   94 & -2152 $\pm$   55 & -2363 $\pm$   78 &DDDD\\
HD 94660         &2453062.7716   & 1.0 & 20 (8)    & 1875 & -2429 $\pm$   70 & -2026 $\pm$   54 & -2204 $\pm$   40 & -2535 $\pm$   71 &DDDD\\
HD 101412        &2453062.7977   & 0.5 & 1350 (4)  & 1575 &   512 $\pm$  111 & -527 $\pm$  313 &   446 $\pm$  106 &   520 $\pm$  129 &DddD\\
HD 141569        &2453062.8424   & 0.5 & 1295 (8)  & 5650 &   -30 $\pm$   40 &    99 $\pm$  123 &   -15 $\pm$   37 &   -94 $\pm$   45 &nnnn\\
HD 144432        &2453062.8983   & 0.5 & 2100 (26) & 4575 &   -30 $\pm$   44 &   -13 $\pm$   27 &    -6 $\pm$   24 &   -71 $\pm$   53 &nnnn\\
HD 76534         &2453063.7617   & 0.5 & 1900 (28) & 4625 &    38 $\pm$   78 &    96 $\pm$   67 &    31 $\pm$   49 &   -29 $\pm$   77 &nnnn\\
HD 104237        &2453063.8129   & 0.5 & 372  (12) & 2375 &    18 $\pm$  101 &    35 $\pm$   58 &    43 $\pm$   49 &   141 $\pm$  141 &nnnn\\
HD 142666        &2453063.8535   & 0.5 & 2800 (12) & 2775 &   -76 $\pm$   50 &    68 $\pm$   50 &    15 $\pm$   40 &   -13 $\pm$   75 &nnnn\\
HD 144668        &2453063.9023   & 0.5 & 729 (30)  & 5700 &    -1 $\pm$   19 &     2 $\pm$   40 &   -11 $\pm$   16 &   -29 $\pm$   23 &nnnn\\
HD 85567         &2453064.7311   & 0.5 & 875 (6)   & 2725 &  -127 $\pm$   68 &    70 $\pm$  130 &  -111 $\pm$   75 &  -172 $\pm$   95 &nnnn\\
HD 94509         &2453064.7650   & 0.5 & 1085 (8)  & 2425 &   110 $\pm$   42 &    -6 $\pm$   28 &    36 $\pm$   22 &   117 $\pm$   49 &dnnd\\
HD 95881         &2453064.7918   & 0.5 & 560 (8)   & 2725 &   -47 $\pm$   47 &   -55 $\pm$  121 &   -20 $\pm$   42 &   -40 $\pm$   53 &nnnn\\
HD 97048         &2453064.8221   & 0.5 & 720 (8)   & 2650 &   -95 $\pm$   51 &  -227 $\pm$  202 &  -109 $\pm$   49 &  -135 $\pm$   63 &nnnn\\
HD 98922         &2453064.8413   & 0.5 & 80  (8)   & 2375 &    12 $\pm$   53 &   -25 $\pm$   82 &    12 $\pm$   44 &   -10 $\pm$   79 &nnnn\\
HD 100546        &2453064.8561   & 0.5 & 104 (8)   & 2500 &   142 $\pm$   74 &   -99 $\pm$  216 &   121 $\pm$   69 &   133 $\pm$   96 &nnnn\\
HD 104237        &2453064.8746   & 0.5 & 146 (10)  & 2800 &   -23 $\pm$   64 &   -39 $\pm$   45 &    -2 $\pm$   35 &     4 $\pm$  103 &nnnn\\
HD 132947        &2453064.9131   & 0.5 & 650 (4)   & 1825 &    89 $\pm$   90 &    10 $\pm$  291 &   100 $\pm$   85 &     4 $\pm$  100 &nnnn\\
HD 94509         &2453110.6666  & 0.5 & 1860 (16)& 3175    &   -75 $\pm$   32 &    12 $\pm$   20 &    -10 $\pm$   17 &  -113 $\pm$   36 &  nnnd\\
HD 94509         &2453110.7167  & 0.5 & 1840 (16) & 3300    &    47 $\pm$   31 &    23 $\pm$   20 &     24 $\pm$   16 &    19 $\pm$   36 &  nnnn\\
HD 96441         &2453112.5457 & 0.5 & 192 (14) & 3575    &  -108 $\pm$   43 &    78 $\pm$   65 &    -48 $\pm$   36 &  -102 $\pm$   48 &  nnnn \\ 
HD 97048         &2453112.5858  & 0.5 & 1260 (14) & 3350    &   103 $\pm$   60 &   181 $\pm$  172 &    137 $\pm$   57 &    76 $\pm$   66 &  nndn\\
HD 97048         &2453112.9974  & 0.5 & 2160 (24) & 4250    &    30 $\pm$   44 &   -98 $\pm$  146 &     24 $\pm$   42 &    74 $\pm$   50 &  nnnn\\
HD 97048         &2453117.0020  & 0.5 & 1185 (16) & 3425    &   -99 $\pm$   49 &   -16 $\pm$  173 &    -88 $\pm$   47 &  -129 $\pm$   56 &  nnnn\\
HD 190073        &2453330.5153   & 0.8 & 215 (8)   & 3700 &    82 $\pm$   62 &   -58 $\pm$  107 &    46 $\pm$   38 &    52 $\pm$   90 &nnnn\\
HD 190073$^*$    &2453330.5299   & 0.8 & 320 (8)   & 3200 &    70 $\pm$   65 &    -24 $\pm$  33 &    -7 $\pm$   29 &       ---        &nnn-\\
CPD-53 295       &2453330.5537   & 0.8 & 1750 (8)  & 2925 &   -36 $\pm$   71 &    76 $\pm$   49 &     6 $\pm$   36 &    -3 $\pm$   81 &nnnn\\
CPD-53 295$^*$   &2453330.5829   & 0.8 & 1200 (4)  & 1800 &    57 $\pm$   95 & 139 $\pm$ 35      &   129 $\pm$32    &  ---    &    nDDn\\  
HD 293782        &2453330.6130   & 0.8 & 1950 (8)  & 3000 &    16 $\pm$   63 &   187 $\pm$   90 &     6 $\pm$   47 &    14 $\pm$   84 &nnnn\\
HD 34282         &2453330.6455   & 0.8 & 1170 (8)  & 3450 &    -8 $\pm$   56 &   124 $\pm$  151 &    28 $\pm$   49 &   -38 $\pm$   68 &nnnn\\
HD 278937        &2453330.6795   & 0.8 & 1860 (8)  & 2550 &    79 $\pm$   112 &   126 $\pm$   83 &    -38 $\pm$  133 &     49 $\pm$ 123    &nnnn\\
HD 275877        &2453330.7110   & 0.8 & 1200 (8)  & 3975 &    80 $\pm$   42 &   -62 $\pm$   59 &    48 $\pm$   32 &   132 $\pm$   50 &nnnd\\
HD 275877B       &2453330.7348   & 0.8 & 1200 (8)  & 1350 &  -283 $\pm$  116 &  -355 $\pm$  195 &  -284 $\pm$   90 &  -400 $\pm$  141 &nndn\\
V380 Ori         &2453330.7576   & 0.8 & 480 (4)   & 1450 &   -38 $\pm$   66 &    44 $\pm$   24 &    34 $\pm$   22 &   -55 $\pm$  391 &nnnn\\
BF Ori           &2453330.7761   & 0.8 & 720 (8)   & 2900 &  -180 $\pm$   38 &   -95 $\pm$   30 &  -144 $\pm$   21 &  -158 $\pm$   45 &dddd\\
HD 287841        &2453330.7985   & 0.8 & 800 (8)   & 2800 &     0 $\pm$   65 &   -66 $\pm$   67 &   -53 $\pm$   42 &   -55 $\pm$   73 &nnnn\\
HD 37411         &2453330.8195   & 0.8 & 720 (8)   & 3250 &   -47 $\pm$   47 &  -498 $\pm$  173 &   -99 $\pm$   46 &     5 $\pm$   57 &nnnn\\
HD 53367         &2453330.8383   & 0.5 & 120 (8)   & 3275 &   143 $\pm$   91 &    13 $\pm$   88 &   102 $\pm$   59 &   275 $\pm$  107 &nnnn\\
HD 53367B        &2453330.8497   & 0.5 & 120 (8)   & 2800 &  -129 $\pm$   94 &  -150 $\pm$   81 &  -117 $\pm$   58 &  -131 $\pm$  113 &nnnn\\
HD 96042         &2453330.8671   & 0.8 & 400 (8)   & 4150 &   118 $\pm$   73 &   -38 $\pm$   76 &    22 $\pm$   52 &    86 $\pm$   82 &nnnn\\
TY Cra           &2453331.5287   & 0.8 &1160 (8)   & 2925 &    14 $\pm$   95 &  -423 $\pm$  229 &   -70 $\pm$   88 &  -465 $\pm$  200 &nnnn\\
HD 17081         &2453331.5530   & 0.5 & 12  (8)   & 3800 &   113 $\pm$   41 &  -219 $\pm$   93 &    38 $\pm$   39 &   135 $\pm$   45 &nnnn\\
HD 17081$^*$     &2453331.5617   & 0.5 & 20  (8)   & 3350 &    64 $\pm$  60  &   -8 $\pm$ 51      &    19 $\pm$ 39    &  ---    &    nnn-\\   
HD 35929         &2453331.5940   & 0.8 & 690 (8)   & 4725 &     3 $\pm$   58 &   -27 $\pm$   33 &   -21 $\pm$   26 &    20 $\pm$   70 &nnnn\\
HD 244604        &2453331.6217   & 0.8 & 1600 (8)  & 4150 &     1 $\pm$   52 &    14 $\pm$   75 &    25 $\pm$   39 &    39 $\pm$   58 &nnnn\\
HD 245185        &2453331.6527   & 0.8 & 1600 (8)  & 3925 &    51 $\pm$   57 &   -37 $\pm$  121 &    20 $\pm$   51 &    68 $\pm$   61 &nnnn\\
HD 31293         &2453331.6774   & 0.8 & 140 (8)   & 4050 &    25 $\pm$   37 &  -290 $\pm$  142 &   -18 $\pm$   35 &    19 $\pm$   67 &nnnn\\
HD 31648         &2453331.6929   & 0.8 & 319 (8)   & 4225 &    54 $\pm$   36 &   127 $\pm$   78 &    63 $\pm$   31 &   -50 $\pm$   53 &nnnn\\
HD 36112         &2453331.7097   & 0.8 & 320 (8)   & 3725 &   -69 $\pm$   68 &  -166 $\pm$   56 &  -149 $\pm$   41 &   -31 $\pm$   77 &nddn\\
HD 35187         &2453331.7310   & 0.5 & 480 (8)   & 3550 &   -53 $\pm$   48 &   -94 $\pm$   76 &   -55 $\pm$   37 &   -33 $\pm$   53 &nnnn\\
HD 35187B        &2453331.7502   & 0.5 & 480 (8)   & 2900 &   -19 $\pm$   61 &    99 $\pm$   52 &    12 $\pm$   36 &   -29 $\pm$   71 &nnnn\\
NX Pup           &2453331.7740   & 0.8 & 1180 (8)  & 4075 &   -26 $\pm$   40 &   172 $\pm$   80 &     2 $\pm$   31 &   -71 $\pm$   46 &nnnn\\
HD 52721         &2453331.7960   & 0.8 & 80  (8)   & 3625 &  -155 $\pm$  154 &   247 $\pm$  149 &   112 $\pm$   91 &  -155 $\pm$  154 &nnnn\\
HD 53179         &2453331.8120   & 0.8 & 320 (8)   & 1875 &   -11 $\pm$   31 &  -101 $\pm$   54 &   -32 $\pm$   19 &  -126 $\pm$   66 &nnnn\\
HD 87403         &2453331.8377   & 0.8 & 1860 (8)  & 3900 &   -36 $\pm$   38 &   244 $\pm$  132 &     9 $\pm$   36 &   -49 $\pm$   42 &nnnn\\
HD 97048         &2453331.8695   & 0.8 & 960 (8)   & 3925 &   -11 $\pm$   53 &  -232 $\pm$  229 &   -37 $\pm$   50 &   -54 $\pm$   57 &nnnn\\
Ty Cra$^*$       &2453332.5265   & 0.8 & 1440 (8)  &  2400&   172 $\pm$ 140    & 28 $\pm$ 134      &  117 $\pm$ 82    & --- & nnn-   \\ 
T Ori            &2453332.6371   & 0.8 & 1440 (8)  & 1900 &   -20 $\pm$   70 &   246 $\pm$  217 &    21 $\pm$   66 &   -32 $\pm$   84 &nnnn\\
HD 37258         &2453332.6568   & 0.8 & 768 (8)   & 2050 &    12 $\pm$   76 &    63 $\pm$  131 &     6 $\pm$   62 &   -82 $\pm$   88 &nnnn\\
HD 37357         &2453332.6714   & 0.8 & 328 (8)   & 2100 &   -21 $\pm$   78 &   -35 $\pm$  163 &     5 $\pm$   69 &   -31 $\pm$   88 &nnnn\\
HD 37490         &2453332.6798   & 0.5 & 8   (8)   & 1950 &   243 $\pm$  162 &   -41 $\pm$  151 &   -31 $\pm$  102 &   243 $\pm$  162 &nnnn\\
HD 37490$^*$     &2453332.6925   & 0.5 & 15  (8)   & 3225 &    34 $\pm$  160    &  -90 $\pm$ 75      &   -49 $\pm$ 60    &    ---     &nnn-\\ 
HD 37806         &2453332.7019   & 0.8 & 144 (8)   & 2125 &   133 $\pm$   58 &  -145 $\pm$  101 &    87 $\pm$   46 &   151 $\pm$   66 &nnnn\\
HD 37806$^*$     &2453332.7200   & 0.8 & 221 (8)   & 3025 &    90 $\pm$ 33    &   6 $\pm$ 37      &    55 $\pm$ 25    &   ---   & nnn-   \\ 
HD 250550        &2453332.7373   & 0.8 & 480 (8)   & 1850 &    56 $\pm$   51 &    41 $\pm$  120 &    52 $\pm$   47 &    69 $\pm$   96 &nnnn\\
HD 259431        &2453332.7562   & 0.8 & 320 (8)   & 2025 &    20 $\pm$   73 &    -5 $\pm$  118 &    24 $\pm$   65 &  -128 $\pm$  179 &nnnn\\
HD 68695         &2453332.7761   & 0.8 & 640 (8)   & 1950 &  -161 $\pm$   99 &   -41 $\pm$  214 &   -83 $\pm$   80 &  -187 $\pm$  107 &nnnn\\
HD 72106         &2453332.7957   & 0.5 & 326 (8)   & 1975 &   162 $\pm$   70 &   -11 $\pm$   91 &    94 $\pm$   55 &   177 $\pm$   78 &nnnn\\
HD 72106B        &2453332.8138   & 0.5 & 326 *8)   & 1750 &    52 $\pm$   90 &     3 $\pm$  122 &    76 $\pm$   67 &    41 $\pm$   95 &nnnn\\
HD 104237        &2453332.8317   & 0.5 & 96  (8)   & 2400 &  -188 $\pm$   85 &   -50 $\pm$   57 &   -89 $\pm$   44 &  -186 $\pm$  184 &nnnn\\
HD 100453        &2453332.8465   & 0.8 & 120 (8)   & 2000 &   111 $\pm$   73 &   -94 $\pm$   77 &   -32 $\pm$   52 &   103 $\pm$   79 &nnnn\\
HD 94660$^*$     &2453332.8614   & 0.5 & 96 (16)   &  3525&     -2337 $\pm$ 40    &  -2144 $\pm$ 12      &   -2159 $\pm$ 11    &  ---    & DDD-   \\ 
HD 94660         &2453332.8738   & 0.5 & 40 (8)    & 2425 & -2437 $\pm$   51 & -2074 $\pm$   44 & -2246 $\pm$   32 & -2406 $\pm$   58 &DDDD\\
\hline
\hline
\end{longtable}
\end{small}

\end{document}